# Analysis of RF losses and material characterization of samples removed from a $Nb_3Sn$-coated superconducting RF cavity

Uttar Pudasaini[1], Grigory V. Eremeev[2], Charles E. Reece[2], James Tuggle[3], and Michael J. Kelley[1,2,3]
[1]Applied Science Department, College of William and Mary, Williamsburg, VA 23188 USA
[2]Thomas Jefferson National Accelerator Facility, Newport News, VA 23606, USA
[3]Virginia Polytechnic Institute and State University, Blacksburg, VA 24061, USA

## *Abstract*

$Nb_3Sn$ ($T_c \approx 18$ K and $H_{sh} \approx 400$ mT) is a prospective material to replace Nb ($T_c \approx 9$ K and $H_{sh} \approx 200$ mT) in SRF accelerator cavities for significant cost reduction and performance enhancement. Because of its material properties, $Nb_3Sn$ is best employed as a thin film (coating) inside an already built RF cavity structure. A particular test cavity noted as C3C4 was a 1.5 GHz single-cell Nb cavity, coated with $Nb_3Sn$ using Sn vapor diffusion process at Jefferson Lab. Cold measurements of the coated cavity indicated the superconducting transition temperature of about 18 K. Subsequent RF measurements indicated field-dependent surface resistance both at 4.3 K and 2.0 K. After initial cold measurements, the cavity RF loss distribution was studied with a thermometry mapping system. Loss regions were identified with thermometry and were cut out for material analysis. The presence of significantly thin patchy regions and other carbon-rich defects is associated with strong local field-dependent surface resistance. This paper summarizes RF and thermometry results correlated with material science findings.

## I. INTRODUCTION

$Nb_3Sn$ is one of the materials that has the potential to replace niobium in some applications of superconducting accelerating structures. It has a critical temperature, a BCS-like energy gap, and a predicted superheating field nearly twice that of niobium [1]. Accordingly, $Nb_3Sn$ promises an RF critical field twice that of niobium and orders of magnitude lower surface resistance. The research into potential applications of this compound started almost from its discovery by Bernd Matthias in 1954 [2]. Its application to the superconducting RF field was also pursued by several groups around the world [3-10]. In the course of these research efforts, many R&D cavities have been coated and tested. The best cavities reached above $E_{acc} > 15$ MV/m and demonstrated quality factors above $10^{11}$ in the superfluid helium bath at 2.0 K.

The $Nb_3Sn$-coated cavities seem to be limited not by the intrinsic RF properties of $Nb_3Sn$ superconductor, but by "extrinsic" factors such as localized defects in $Nb_3Sn$ coatings. Research efforts are presently focused on the characterization of such defects, understanding of their formation, and eventually their reduction or elimination from $Nb_3Sn$ coatings during film growth and subsequent processing, e.g. see [11-13]. In this study, we investigated lossy regions that were identified using the JLab thermometry system during SRF cavity testing. The lossy regions of interest were cut out of the cavity after the test and inspected with surface science characterization techniques. Besides features known as "patches," which have been commonly found and discussed, other surface features, contributing to losses, were identified. The identified coating structure and surface features were correlated to RF losses.

## II. CAVITY COATING

"C3C4" was a 1.5 GHz single cell cavity made from high purity (≈ 300) fine-grain Nb. It was subjected to buffered chemical polishing (BCP) using a solution of 49% HF, 70% $HNO_3$, and 85% $H_3PO_4$ in the ratio of 1:1:2 for removal of about 20 $\mu$m inside and 5 $\mu$m outside. It then received high pressure rinsing (HPR) with ultra-pure water, assembled, and evacuated before the baseline test at 2 K. The cavity was limited by high field Q-slope at $E_{acc} \approx$ 27 MV/m [1]. The low field quality factor was about $1.6 \times 10^{10}$ at 10 MV/m. After the baseline test, the cavity was removed from the test stand in the cleanroom, followed by HPR, and allowed to dry in the cleanroom. For the $Nb_3Sn$ coating, 99.9999% Sn shots and 99.99% $SnCl_2$ powder purchased from American Elements were used. Sn and $SnCl_2$ were packaged in niobium foil. Each package contained either 1±.0.1 g of Sn or 0.5±.05 g of $SnCl_2$. Three packages of Sn containing 3 g total and six packages of $SnCl_2$ containing 3 gr total were placed inside the cavity onto the niobium foil covering the bottom flange. The top flange of the cavity was covered with Nb foil as well. Nb foil was commercial grade unalloyed niobium purchased from Eagle Alloys. The cavity was assembled in the cleanroom for coating and double bagged before being transferred to the thin film lab. In the deposition lab, the cavity was attached to the deposition chamber support assembly and installed in the $Nb_3Sn$ coating chamber. The setup was evacuated for a couple of hours, and the heating profile was initiated. In Fig. 1 the temperature profile from the three heat zones of the furnace and the furnace pressure are shown. After the heating run and furnace cooldown, the insert and the furnace were purged with $N_2$ to atmospheric pressure, and the cavity was removed from the insert. A description of the coating system is available in [14]. In Fig. 2 (a) the cavity is shown sitting on the table after removal from the insert. When the Nb cover foils were removed, several features were observed: discoloration was seen on NbTi flanges (b), residues were observed on the bottom Nb cover and the Nb foils that contained Sn and $SnCl_2$ (c), and Sn condensation on the top Nb foil (d).

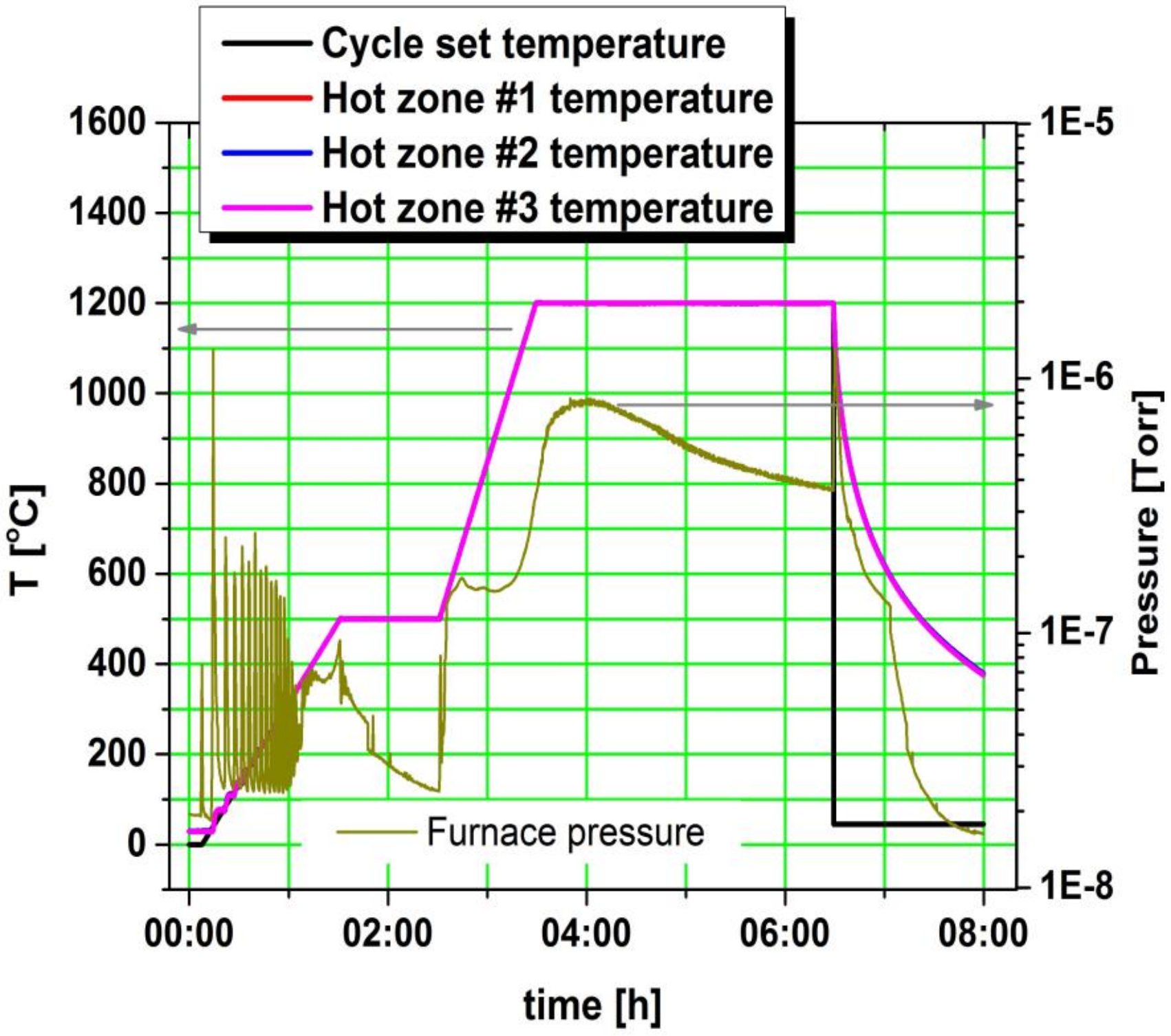

Fig. 1. The temperature and pressure profile from the first $Nb_3Sn$ cavity coating run. Note the three separate regions: first, the temperature is increased at 6 ˚C per minute, then, it is parked at 500 ˚C to allow for $SnCl_2$ evaporation, then, the temperature is increased at 12 ˚C per minute to 1200 ˚C, where the furnace temperature remains for 3 hours, finally, the heat is turned off and the furnace cools down.

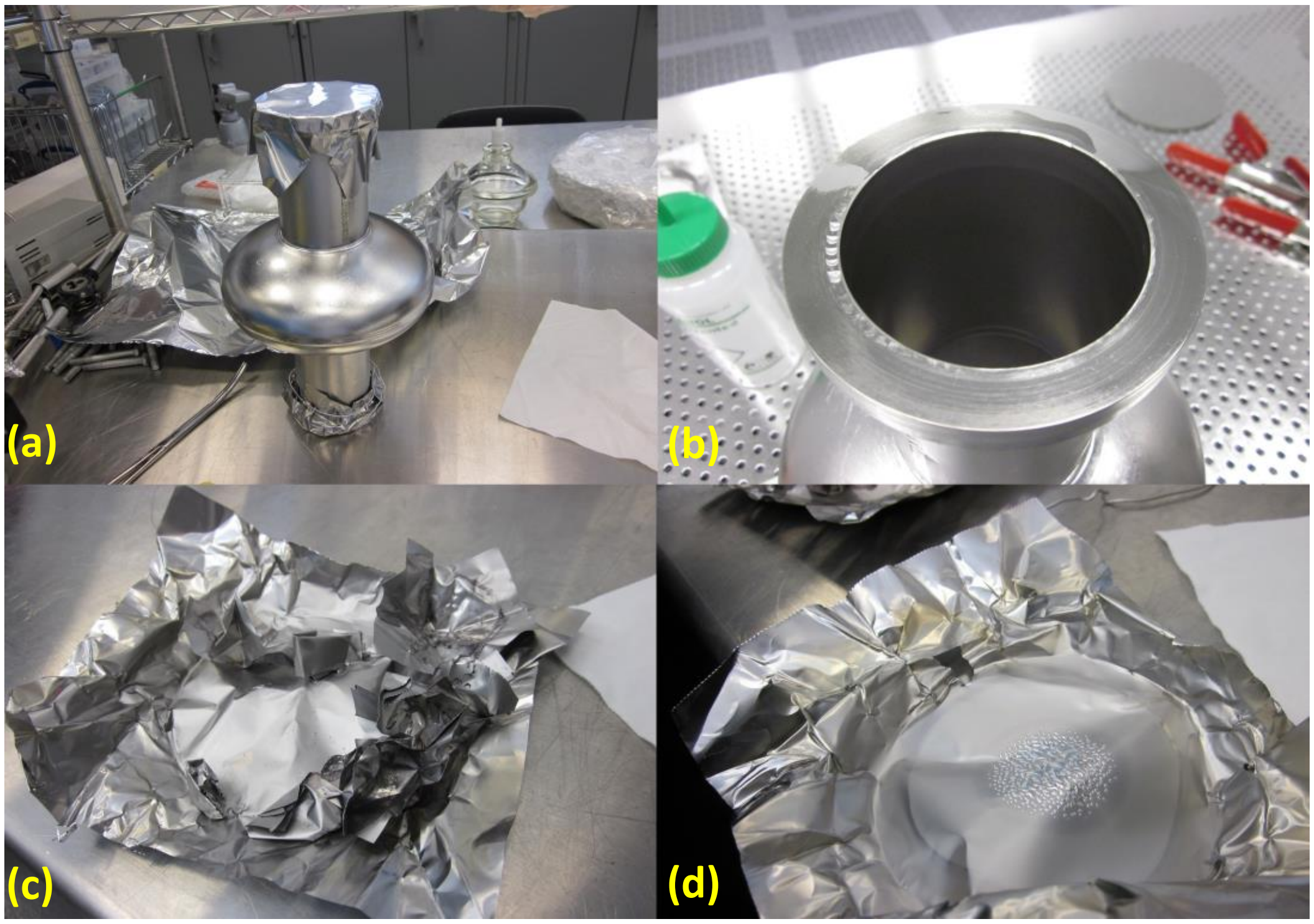

Fig. 2. Cavity C3C4 after the coating run. Note discoloration on the NbTi cavity flange (b) and 'tin' droplets on the niobium foil from the cavity top flange (d).

## III. OPTICAL INSPECTION

After the coating, the cavity interior was optically inspected with a KEK-style optical inspection bench [15]. The optical inspections of C3C4 revealed complete coverage of the surface. In one place at the equator area of C3C4, we found several ≈ 100 $\mu$m-size features, Fig. 3(marked with red ellipses).

## IV. RF TEST RESULTS

After optical inspections, the cavity was RF tested at 4.3 K and 2.0 K. Typically, four Lakeshore DT-670 diodes were attached to the cavity to monitor the temperature spread across the cavity during cooldowns: one diode attached to the bottom beam tube, one to the bottom half cell, one to the top half cell, and the last diode was attached to the top beam tube, 90 degrees apart azimuthally each. A network analyzer was used to monitor the resonance frequency and the quality factor (based on 3 dB measurement) of the cavity during cooldowns. The superconducting transition temperature of C3C4 was found to be 17.9±0.25 K. We did not observe any transition at 9 K, which indicates complete $Nb_3Sn$ coverage on the Nb. After $Nb_3Sn$ coating, the cavity was limited by a strong slope in $Q_0$ at low fields at both 2.0 and 4.3 K helium bath temperature, Fig. 4. The cavity was then re-tested with a temperature mapping array installed around the cavity [17]. The $Q_0$ vs $E_{acc}$ curve was similar, and the cavity was limited again at low gradients due to a strong slope. The low-field $Q_0$ was better in the re-test, which is attributed to a different cooldown condition. Due to the higher low-field quality factor, a "knee" at about $E_{acc}$ = 4 MV/m is more evident than in the first test.

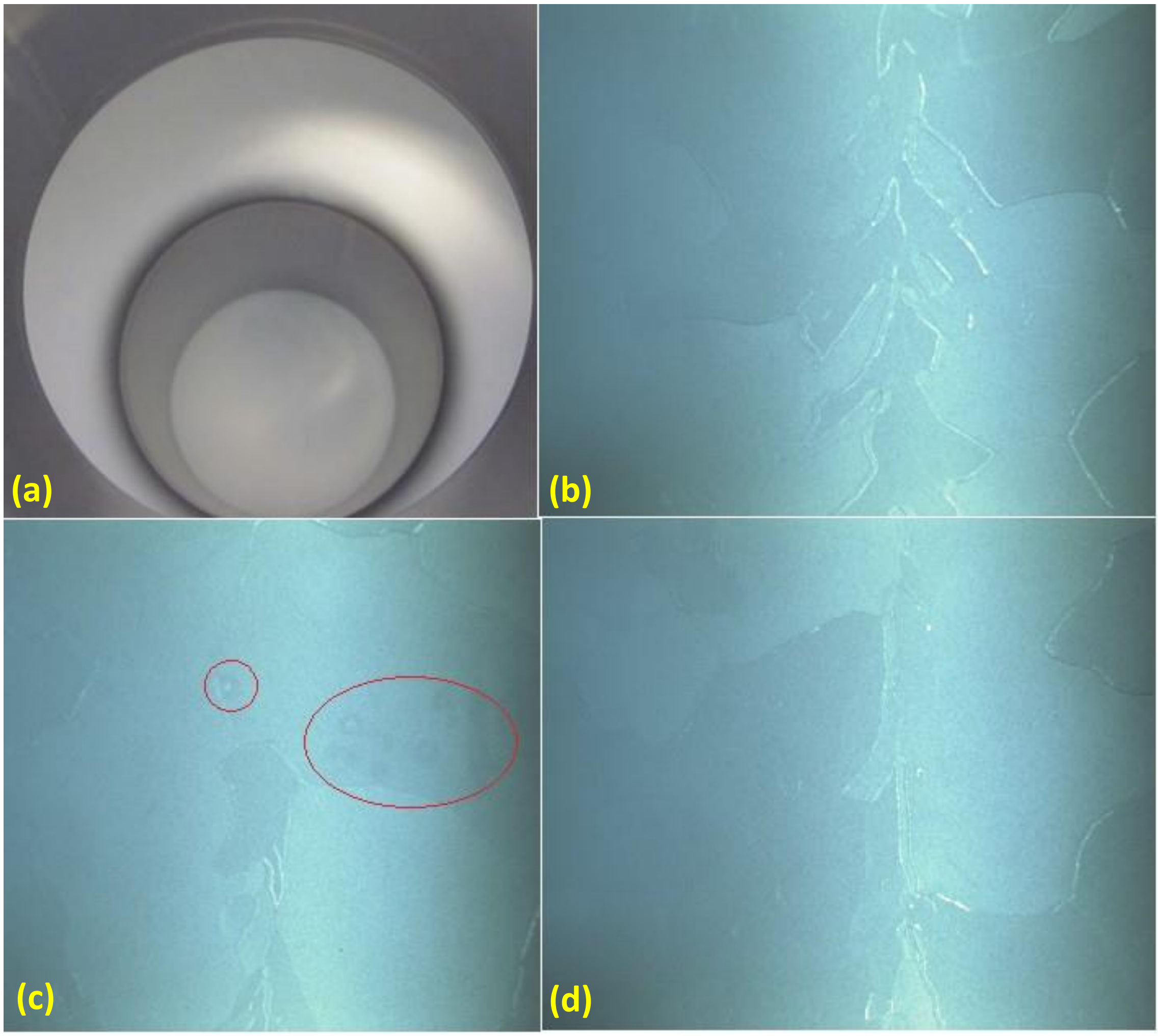

Fig. 3. Optical inspection pictures of C3C4. (a) shows the coated surface of the cavity viewed via one of the beam tubes. (b) and (d) show characteristic equatorial weld regions of the coated cavity. (c) shows an equatorial region with several observed features marked with red ellipses.

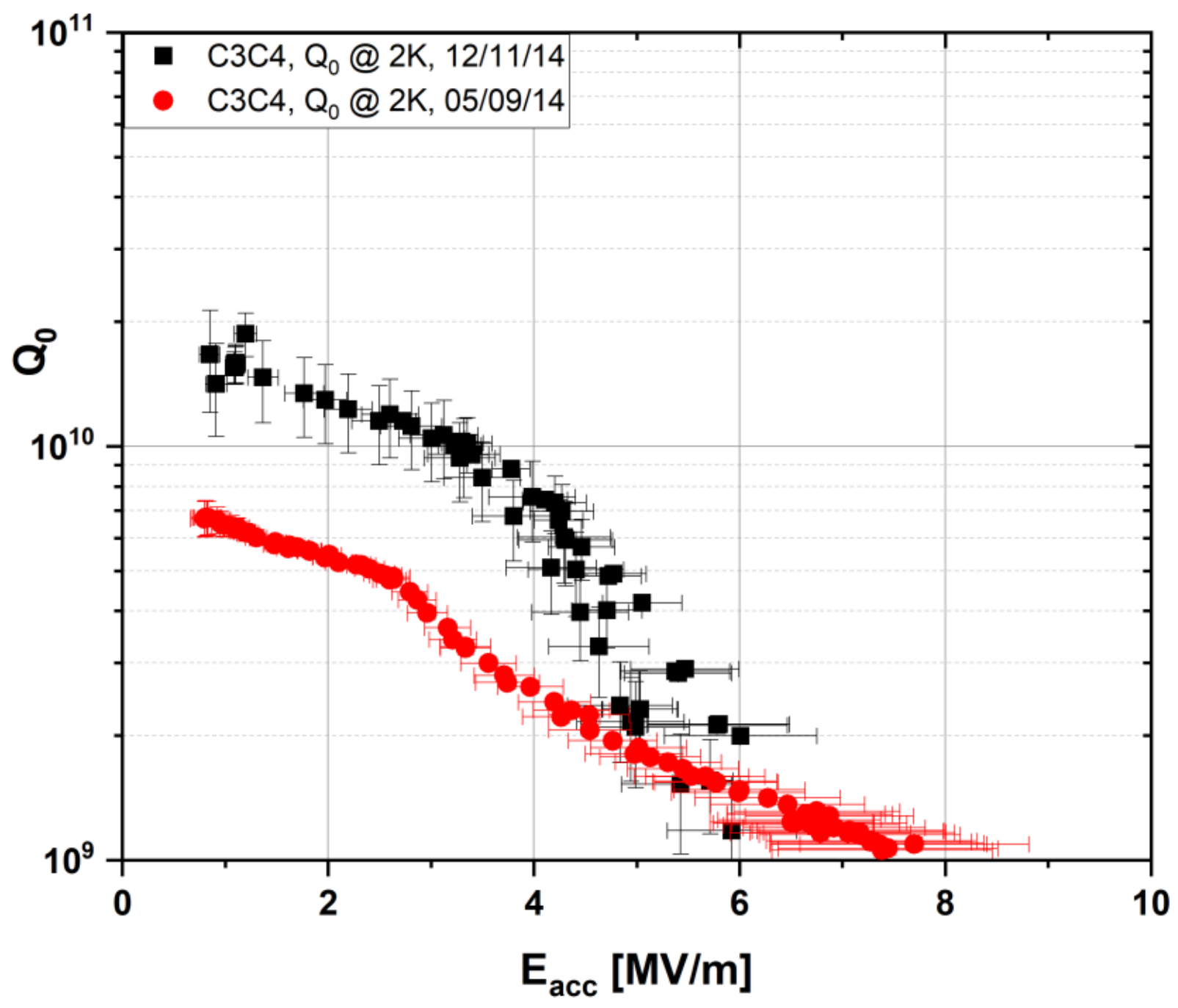

Fig. 4. $Q_0$ vs $E_{acc}$ for cavity C3C4.

During the latest test, temperature maps of the cavity surface were acquired at different field levels with the JLab temperature mapping system [17]. The system is comprised of 36 boards surrounding the cavity at the equally spaced interval. Each board has 16 temperature sensors mounted on spring-loaded pins that push against the outer surface of the cavity. The temperature difference between the outer cavity surface and the Helium bath at the highest gradient showed three regions with the strongest heating on the cavity surface at (15, 3), (7, 8), and (2, 7). The X-axis and Y-axis here represent board number and sensor number respectively. For example, (15, 3) represents the location at board 15 and sensor 3. Sensor 1 is at the top iris, sensor 16 at the bottom iris and sensor 8 is at the equator. To assess the heating in other areas, the maximum of the color scale is reduced to 20 mK in Fig. 5. The three hottest areas mentioned above are now off scale and became white squares. The white square at (26, 5) is an artifact of a malfunctioning temperature sensor. The areas which were cut out are marked with red line. The number next to the marked areas indicated the number assigned to the extracted sample. This temperature map shows that the heating on the surface is not uniform, with some distributed areas, e.g., at (35, 7), showing more heating than others, e.g., (29, 7). This temperature map was sectioned into areas with different heating behavior to guide extraction of characteristic samples from different surface areas.

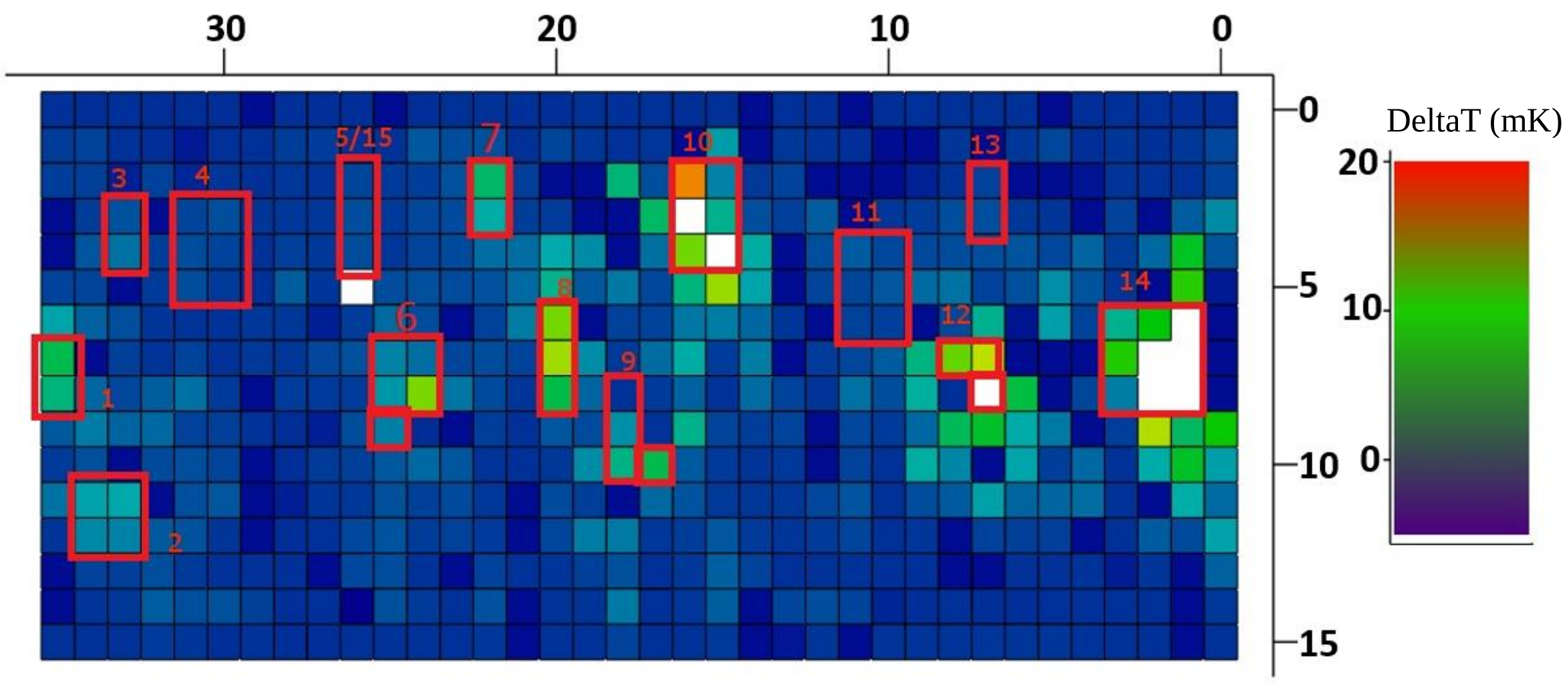

Fig. 5. Temperature map of C3C4 at $E_{acc}$ = 6 MV/m shows the temperature delta due to RF fields on the outer cavity surface. See text for details.

# V. SAMPLE EXTRACTION

After the RF test with temperature mapping, the outside surface of the cavity was marked with an engraver according to the temperature map in Fig. 5. Each area marked for extraction was encircled with engraver roughly following the contour of the thermometers. Several samples were then milled out with a 3/32 two-flute OSG Exomini TiN-coated end mill running at 3200 rpm as illustrated in Fig. 6. The cut was cooled with helium gas flowing through 1/4" tygon tubing. The helium gas nozzle was set approximately 5 cm from the end mill. The temperature of the niobium surface at the cut was checked with an infrared thermometer and never exceeded 40 ˚C during milling. During the milling, attempts were made not to cut through, but to leave a thin layer of material, to reduce contamination and damage to the internal surface from the milling process. This was not always possible due to cavity curvature, but some material was left when the milling was finished. The samples were then removed with pliers and de-burred.

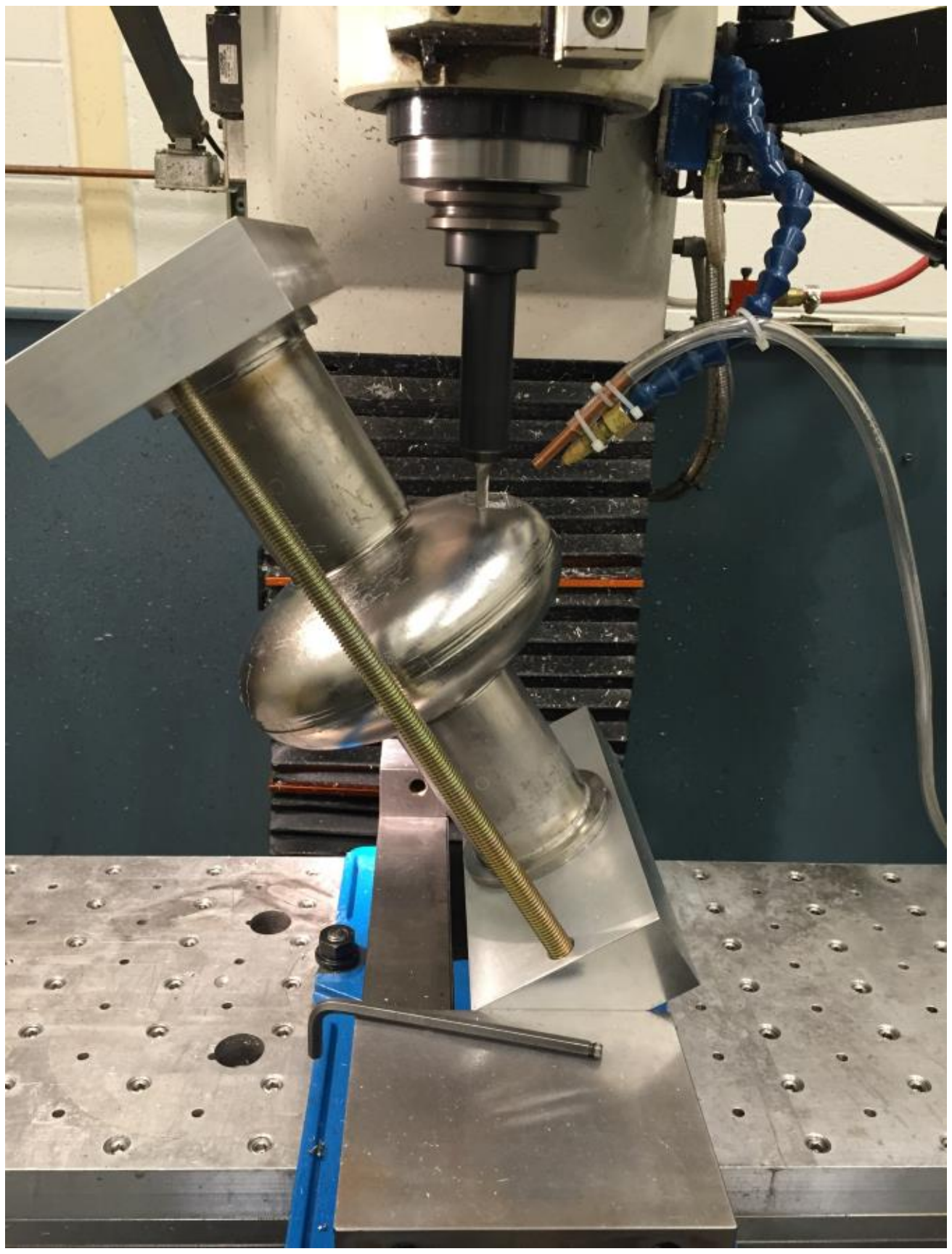

Fig. 6. Setup used to millout marked areas from the cavity C3C4 noted in Fig. 5.

The samples were then rinsed with acetone, then with methanol, and then dried with ionized nitrogen. The samples were then ultrasonically cleaned in a ultra-pure water bath with 2% of micro-90 followed by ultra-pure water rinse. The samples were rinsed again with acetone, then methanol, and finally dried with ionized nitrogen. Figure 7 shows a photo of sample #14 following each cleaning step before surface analysis. The edges of the samples were inspected for cutting artifacts with secondary electron microscopy (SEM). The artifacts included cracks and contamination largely localized near the edge of each cutout. The appearance of cracks and contaminants was consistent with the inference that these features developed during the cutting process. An example of such artifacts is shown in Fig. 8. Cracks were found to have a preferential direction and were intergranular, spreading up to several hundreds of micrometers. These cracks are about 100 nm wide extending to the niobium substrate as shown in Fig. 9. Contamination was mostly the pure niobium that was probably transferred during the cutting process. Traces of Al and Cu, found occasionally, were confirmed to be derived from the tooling used to prepare cutouts. C and O, which are common handling contaminants, were also observed.

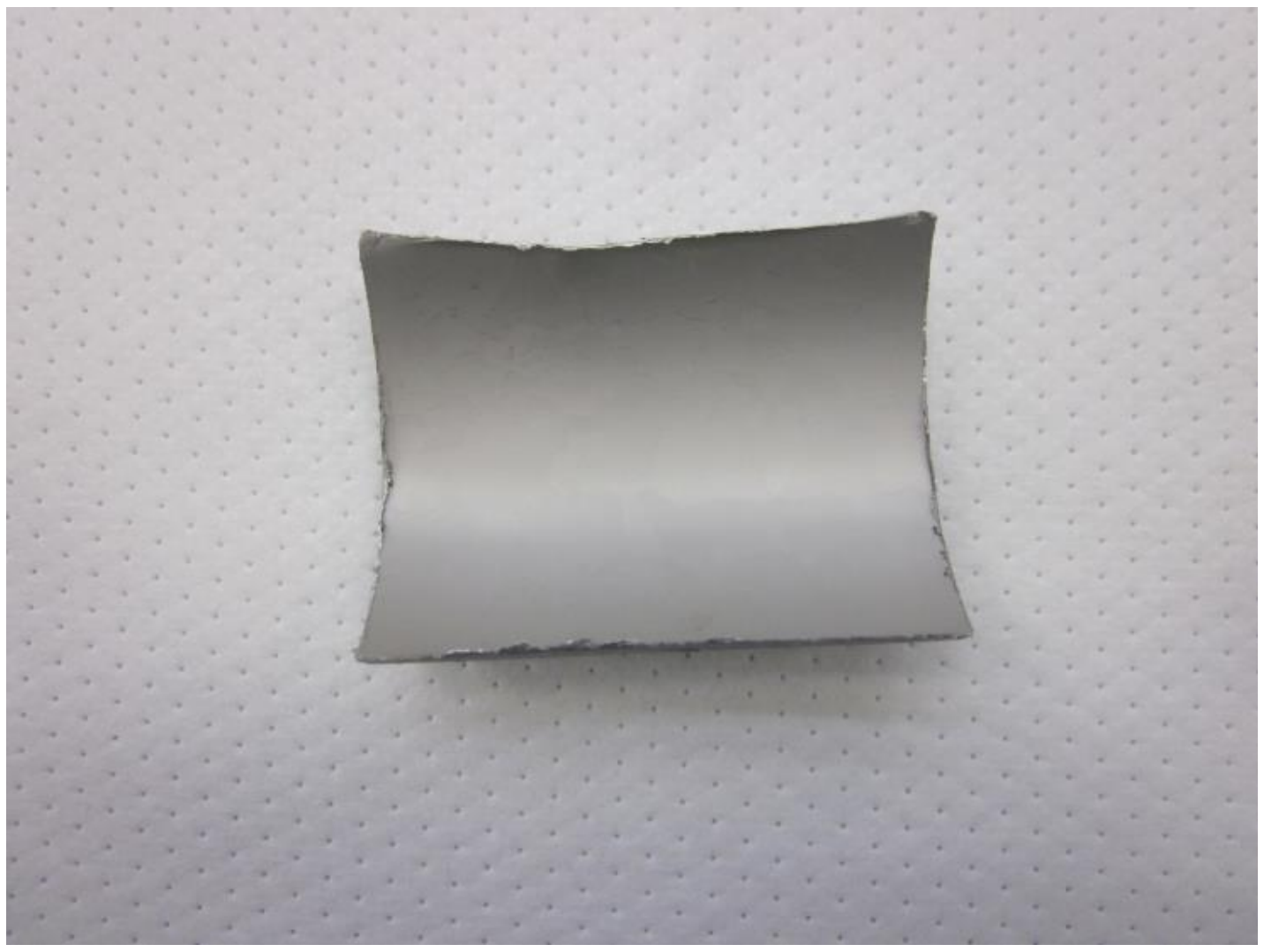

Fig. 7. Sample #14 (CVT14) following each cleaning step before the analysis.

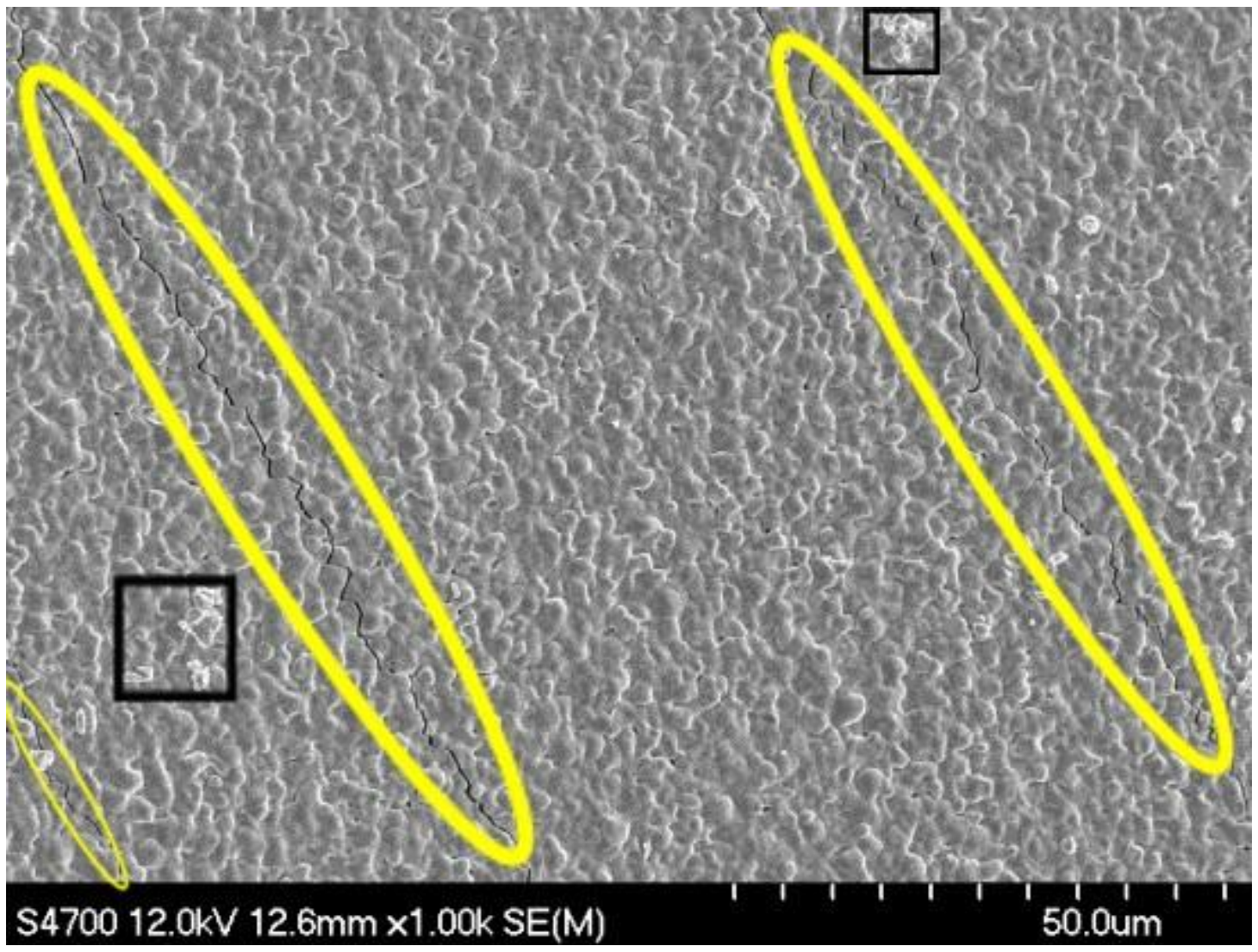

Fig. 8. Cutting artifacts in CVT8. Note that the cracks are aligned in the same direction. Some traces of contamination can be seen, which was found to be niobium.

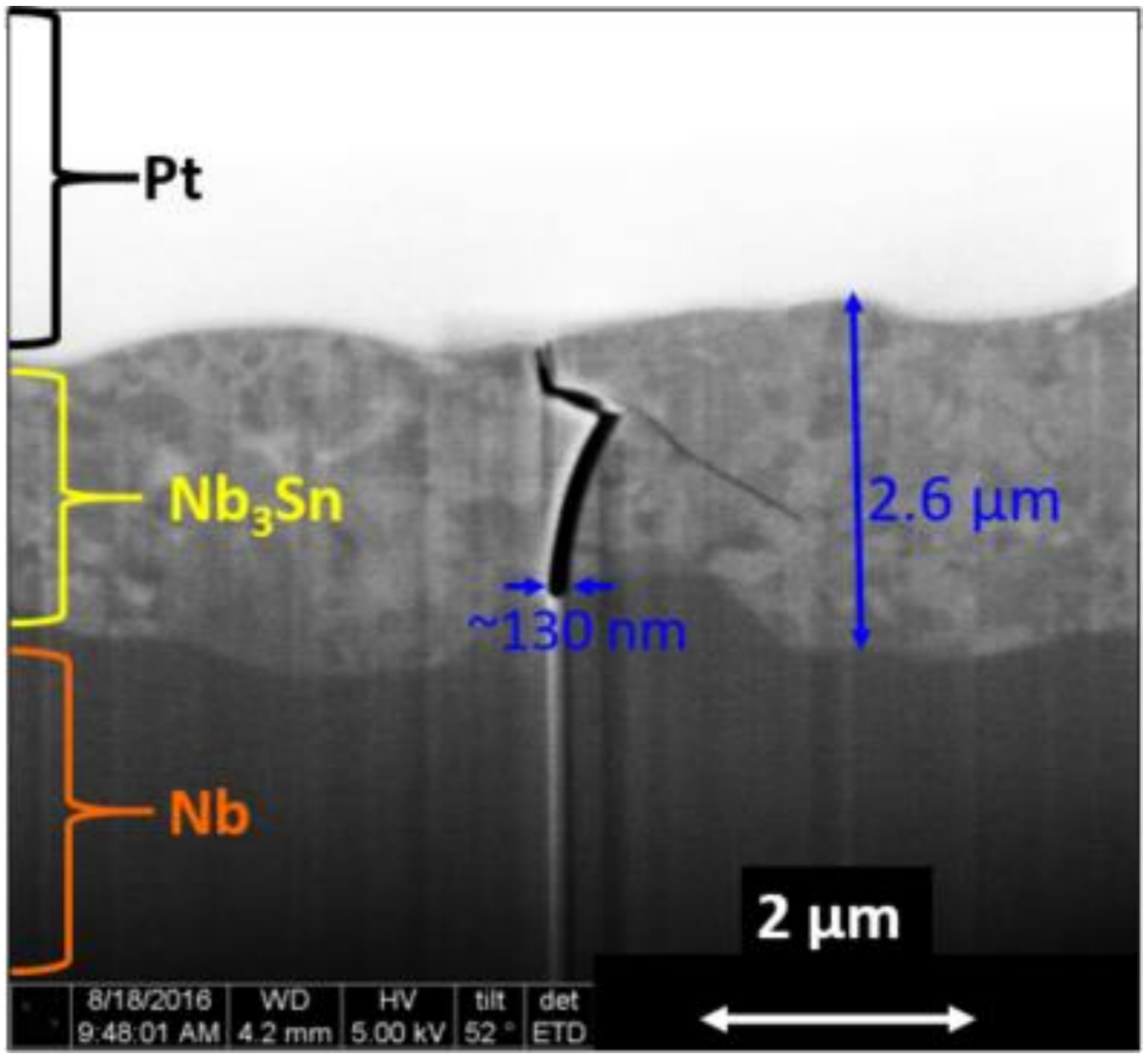

Fig. 9. A cross-sectional view of a crack observed in CVT14. Note that the crack extends down to the Nb-$Nb_3Sn$ interface.

# VI. TEMPERATURE MAPPING DATA ANALYSIS

The typical measurement of the quality factor as a function of field is a measure of the average surface resistance and its field dependence. Temperature mapping during the cryogenic RF test, on the other hand, allows measuring local temperature rise and associate field dependent loss with specific areas of the cavity. Temperature sensors similar to the ones used in this study to measure the surface temperature of the outer cavity surface were measured to be $\eta \cong (35 \pm 13)$ % efficient [18, 19] that is, the temperature delta indicated by the sensor is $\eta$ times smaller than the actual surface temperature. The temperature rise of the outside the cavity surface can be related to the RF power dissipated on the inside via the steady state heat flow equation:

$$\frac{1}{2} R_s H^2 = \frac{1}{R_{Kapitza}(T_s, T_B)} (T_s - T_b) \quad 1$$

where $R_s$ is the RF surface resistance on the RF side of the cavity, H is the surface magnetic field amplitude, $R_{Kapitza}$ is the Kapitza resistance at the helium-cavity exterior interface, $T_s$ is the temperature of the outside surface of the cavity, $T_B$ is the helium bath temperature. Since the exterior surface of C3C4 was not coated with $Nb_3Sn$, Kapitza resistance at the helium niobium interface can be used here. If $\Delta T$ is the temperature rise measured by the temperature sensors, then $T_s - T_b = \frac{\Delta T}{\eta}$ and the previous equation can be re-written:

$$R_s = \frac{2}{R_{Kapitza}\,(T_s, T_B)\eta}\frac{\Delta T}{H^2} \qquad 2$$

The surface resistance can be calculated from the equation above if one knows the Kapitza resistance. Kapitza resistance, however, is very dependent on the material and its preparation. Values of Kapitza resistance at 2 K varying by almost two orders of magnitude have been reported [20-22]. For chemically polished Nb, the typical value $is\ R_{Kapitza} = 1\, cm^2K/W$ [21], which will be used in the analysis with an understanding that this value could be a factor of two or three different for the actual C3C4 surface. While there could be a significant variation in the Kapitza resistance between the cavities, the variation in Kapitza resistance between different areas of C3C4 is expected to be small, since the cavity was always treated as a whole. This assumption implies that, while the surface resistance curves may shift up or down, depending on the Kapitza resistance of the actual surface, the relative position of the curves and the field dependence will be the same.

With the caveat mentioned above, the average RF surface resistance of the cavity can be calculated from the temperature map of the whole cavity surface. Averaging the temperature rise adjusted for the local field amplitude over the entire surface of the cavity, the average surface resistance of the cavity can be calculated from the last equation. The average surface resistance can be calculated this way for each temperature map for each field level. The resulting field dependence of the surface resistance can be converted to the quality factor via $Q_0 = \frac{G}{<R_s>}$, where $G$ = 273 is the geometry factor for this cavity shape. In Fig. 10, the intrinsic quality factor derived this way is shown along with the quality factor measured during RF test using the standard phase-lock loop (PLL) technique. The error bars for the quality factor calculated from the thermometry data are derived from the temperature sensor efficiency spread of 37 % and the sensor noise floor, which was assumed to be 50 $\mu$K. The typical $Q_0$ error bars for PLL RF measurements are about 10 % (not shown in the plot). Considering different areas of the cavity, a difference in the average surface resistance between the top and the bottom half cells can be evaluated. For example, since the Sn source is located at the bottom flange, tin vapor pressure gradient along the cavity axis of symmetry may exist, which may result in a coating gradient. In Fig. 11, the average surface resistance of the top half cell vs. the bottom one calculated using thermometry data as a function of field is shown. The analysis shows the top half cell has a higher surface resistance, but the difference is within the errors of the analysis. The same analysis applied to the cutouts indicated in Fig. 5 is shown in Fig. 12. The temperature rise adjusted for the local field is averaged over the temperature sensors covering the specific cutout area. Figure 12 summarizes the data for all the cutouts.

All cutouts were broken down into three groups by the field dependence of their average surface resistance. The first group consists of the samples that have weak field dependence, that is, the surface resistance stays constant with the applied field up to the highest field. The second group consists of the cutouts, which have a weak field dependence until $E_{acc}$ = 4.5 MV/m, but have a surface resistance switch and a stronger field dependence above this field. The last group of cutouts comprises the surfaces that have a strongly field-dependent surface resistance from the lowest field. In Fig. 13, the average surface resistance for the cutouts that have field independent loss is shown. The surface resistance for these samples stays constant with the field. There is an increase in the surface resistance at the accelerating gradients above about 5 MV/m. The increase is within errors, but is evident for all the thermometers in this group, suggesting a systematic effect. In Fig. 14, the average surface resistance for the cutouts that have a resistance switch at $E_{acc}$ = 4.5 MV/m. There is a jump in the surface resistance at $E_{acc}$ = 4.5 MV/m. This effect is suspected to be caused by a weakly superconducting defect or defects becoming normal conducting at this field level. In Fig. 15, the average surface resistance for the cutouts that have a strong field-dependent loss. The surface resistance for these samples increases exponentially with field at all field levels, and the temperature at the highest field is an order of magnitude higher than the temperature rise measured at the lowest field.

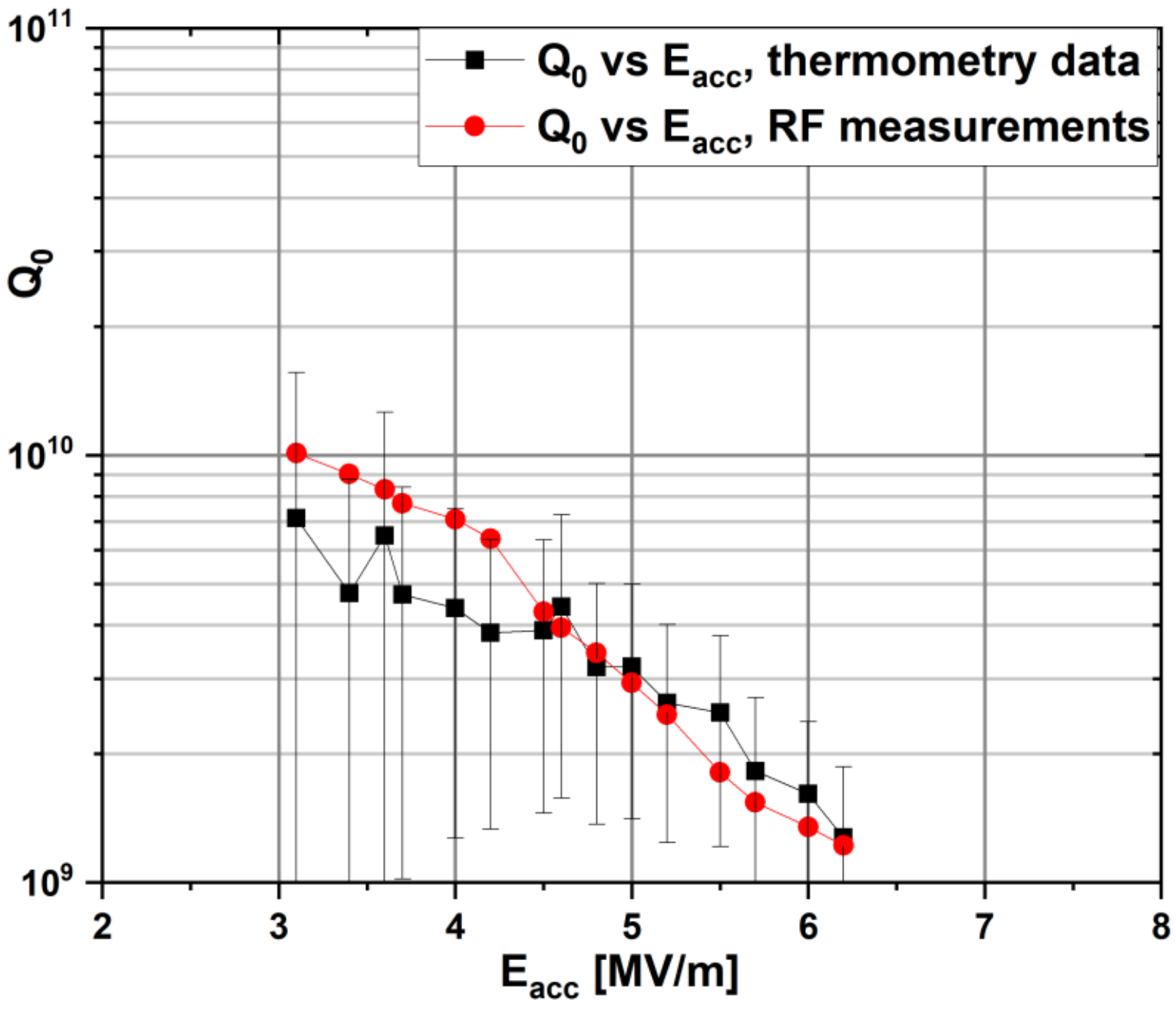

Fig. 10. The intrinsic quality factor derived from both RF and temperature mapping data.

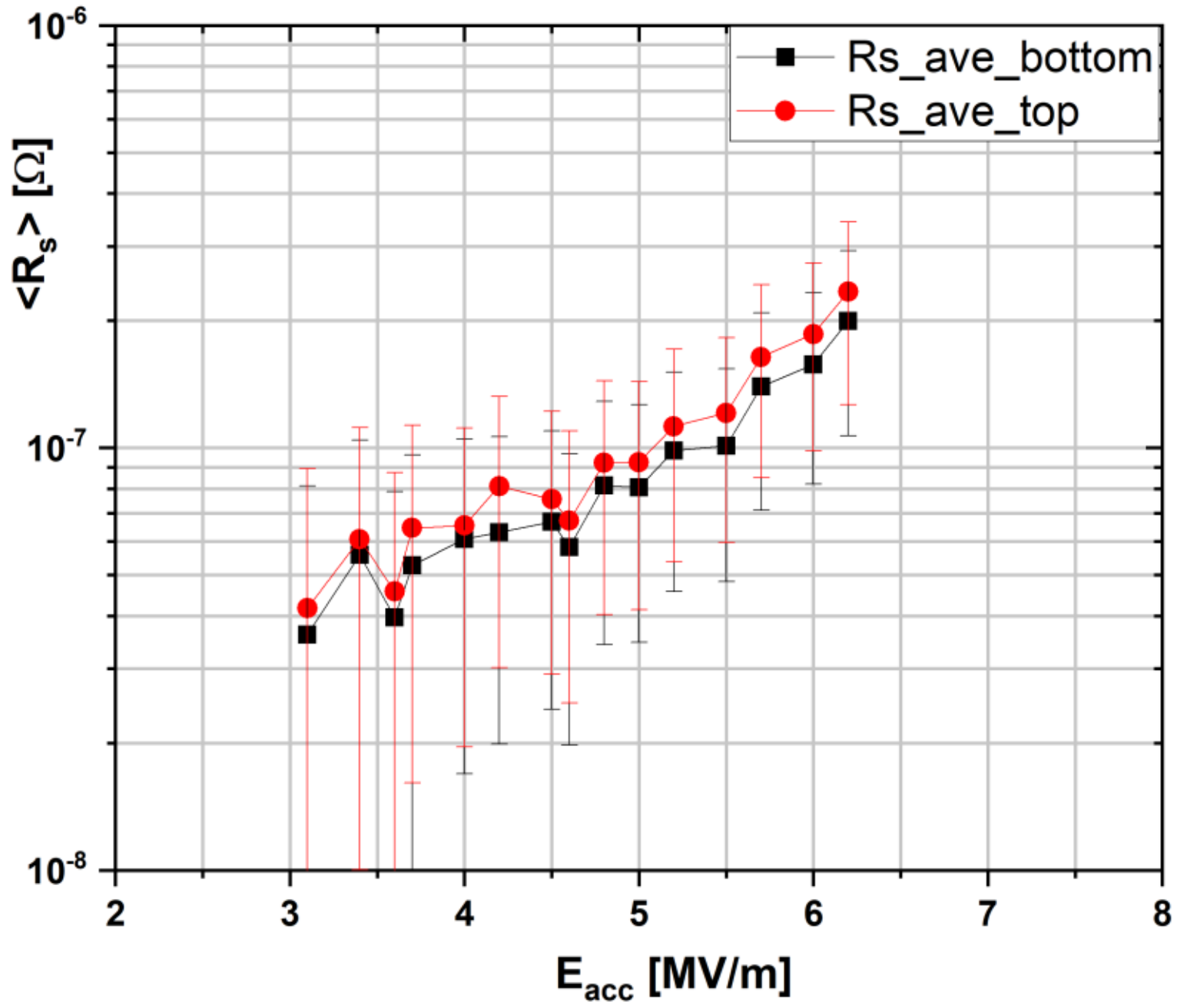

Fig. 11. The average surface resistance of the top and bottom half cells derived from temperature mapping data.

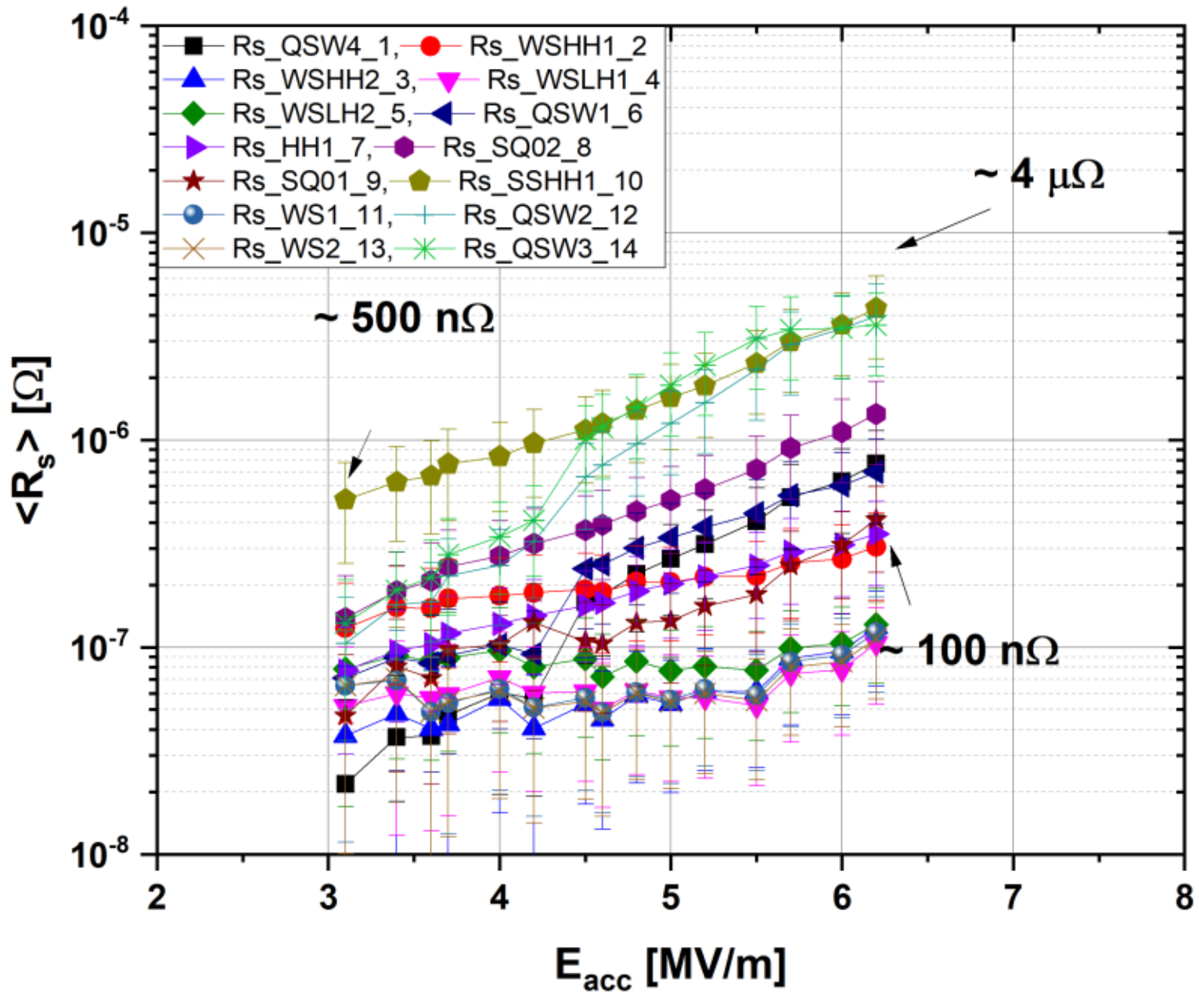

Fig. 12. The calculated surface resistance dependence with field for all cutouts.

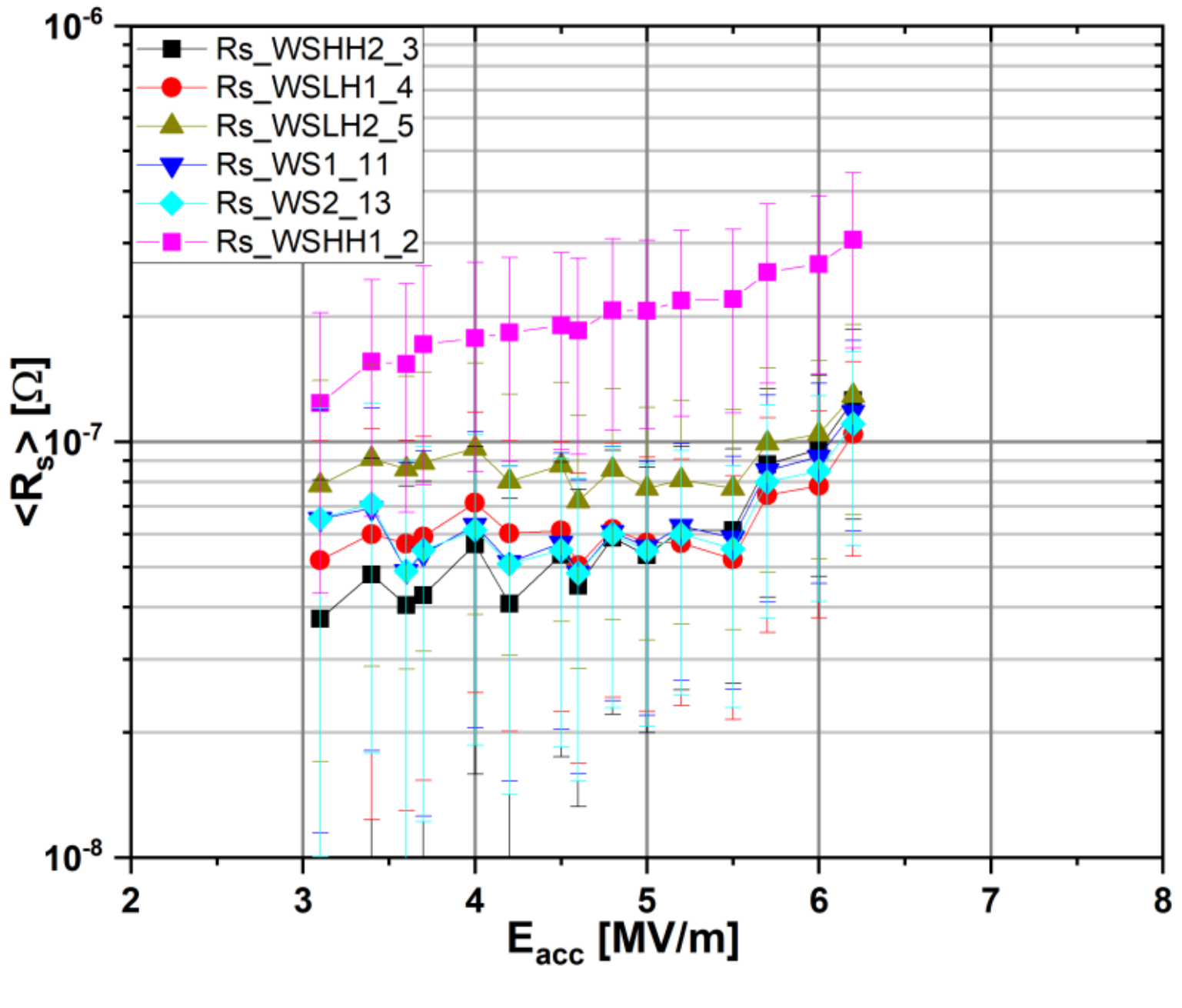

Fig. 13. The average surface resistance for cutouts that show a nearly field-independent loss.

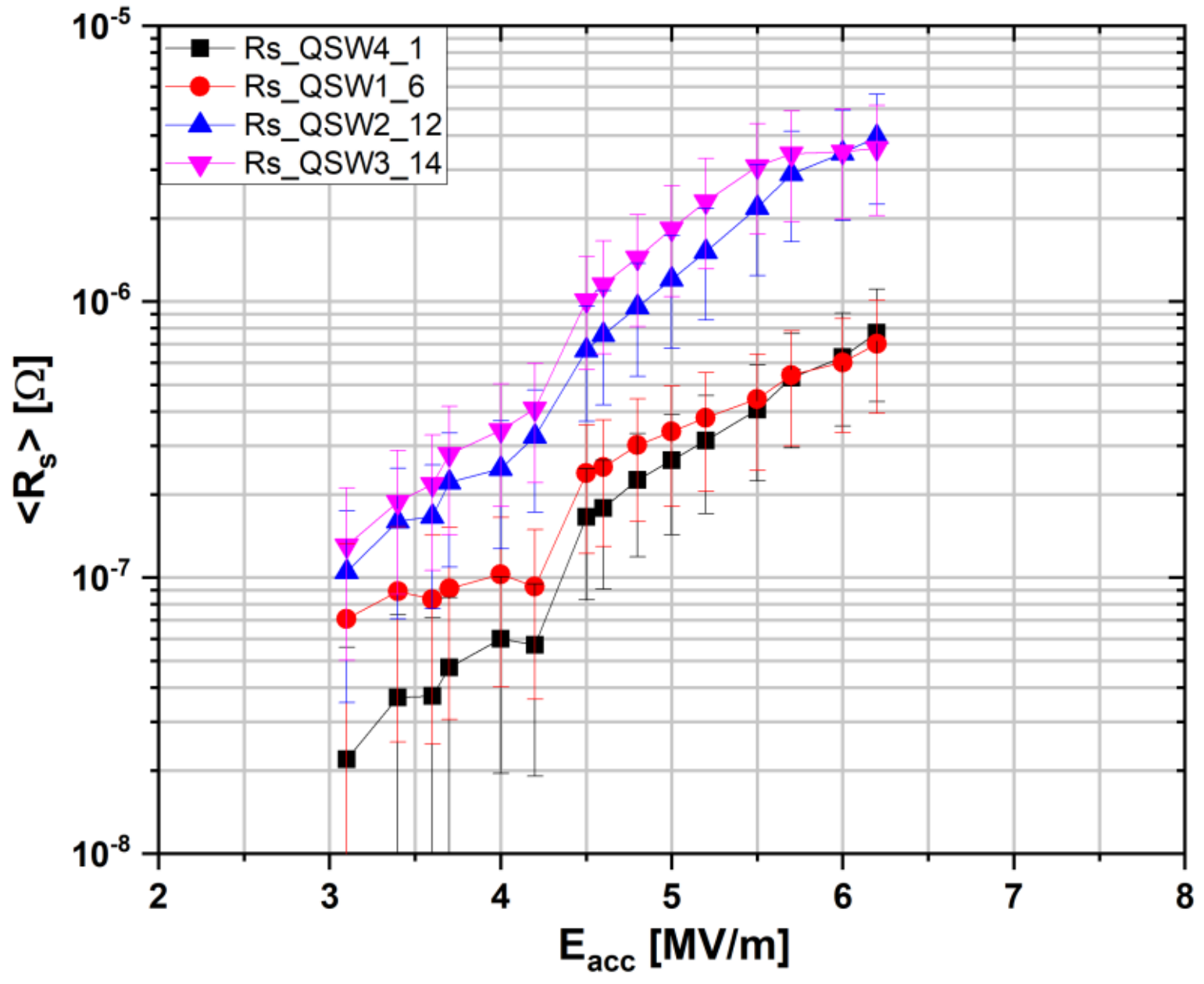

Fig. 14. The average surface resistance for cutouts that have a resistance switch near $E_{acc}$ = 4.5 MV/m.

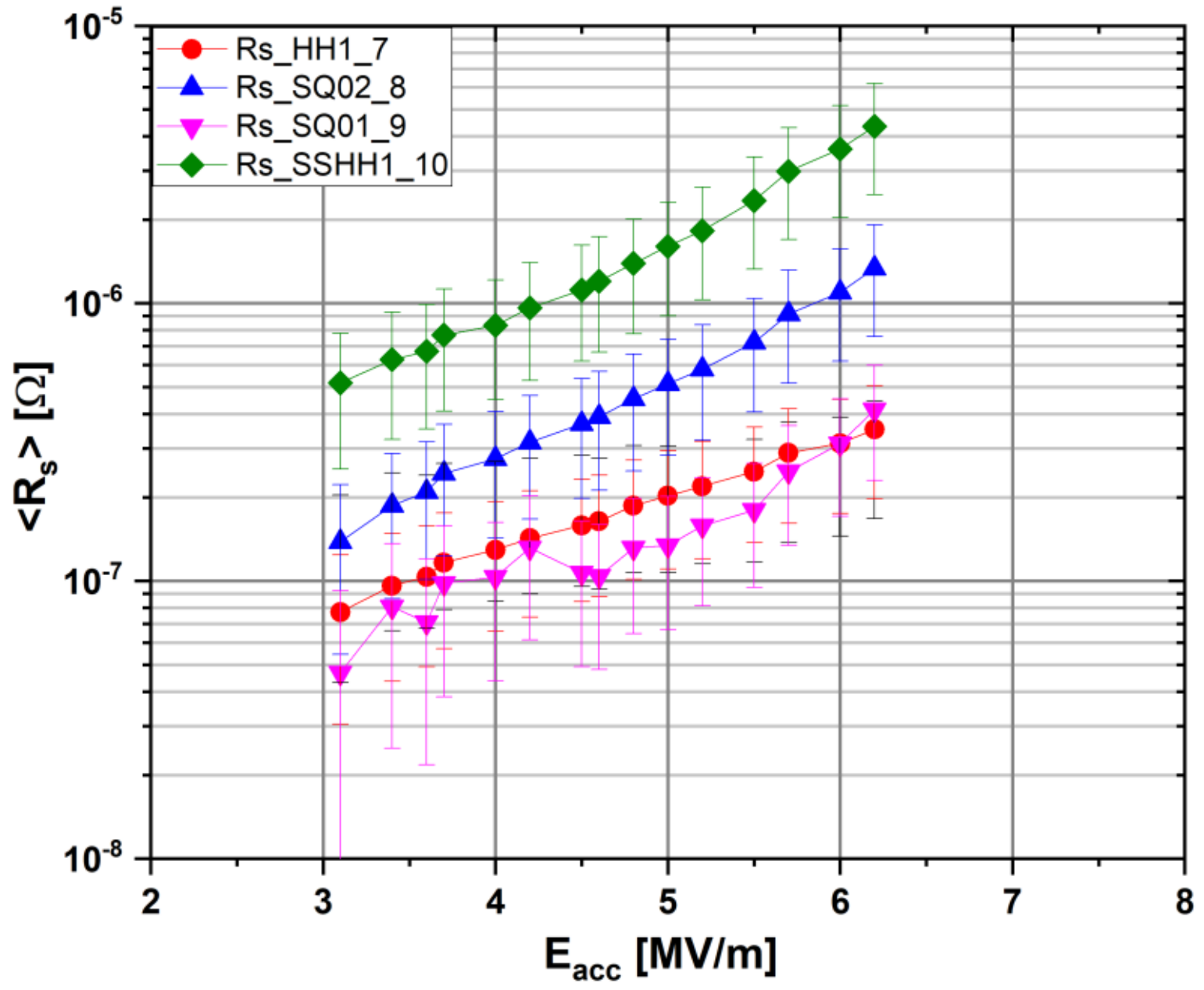

Fig. 15. The average surface resistance for the cutouts that have a strongly consistent field-dependent loss.

# VII. SURFACE ANALYSIS

Six cutouts from the cavity, two samples representing each category of a different trend of surface resistances, and two more cutouts representing the top and bottom beam pipes were analyzed with the surface analytical techniques. They were examined with a field emission scanning electron microscope (FESEM) with energy dispersive X-ray spectroscopy (EDS) for microstructure and local composition. Atomic force microscopy (AFM) was used on some samples to investigate surface topography. Focused ion beam (FIB) cross-sections of cutouts were prepared and examined with electron backscatter diffraction (EBSD) and SEM for the crystal structure and the thickness of the coating. Details on FIB cross-section preparation and EBSD data collection are available in [2].

## *A. Microstructure and Local Composition*

SEM images captured from multiple locations of each sample showed a mostly uniform coating coverage. Observed microstructures were very similar, as shown in Fig. 16. Images at the first, second and third columns belong to different sets of samples that showed field-independent loss (Fig. 13), strong field-dependent loss (Fig. 15), and resistance switch near $E_{acc}$ = 4.5 MV/m (Fig. 14), respectively. The composition was measured at several locations of each sample with EDS at 12 or 15 kV. Average compositions of each sample were found very similar as shown in Table *I*. Average tin content sensed in each cutout was slightly lower than the nominal $Nb_3Sn$.

On the other hand, cavity samples revealed several voids, including microscopic pits in the coating as shown in Fig. 17(a). Note that the term *pit* here refers to the void with well-defined sharp edges at the perimeter. Pit diameters were found to vary between 200 nm to 600 nm. Pits were found normally at vertices where multiple grain boundaries meet as shown in Fig. 17(b) and 17(c) The presence of these voids with sharp edges is understood to be harmful to RF performance of a cavity due to local current and magnetic field enhancement [23, 24]. SEM images with a magnification of 3500 - 5000 were sampled from 15-20 randomly selected areas of cutouts. Voids were counted from those images to get an estimation of void density in each cutout. The estimated average number of voids for all the samples was found to be (11±2) per thousand square micron area, shown in Table *I*. Approximately one third of the observed voids were pits like as shown in Fig.17(b) or 17(c).

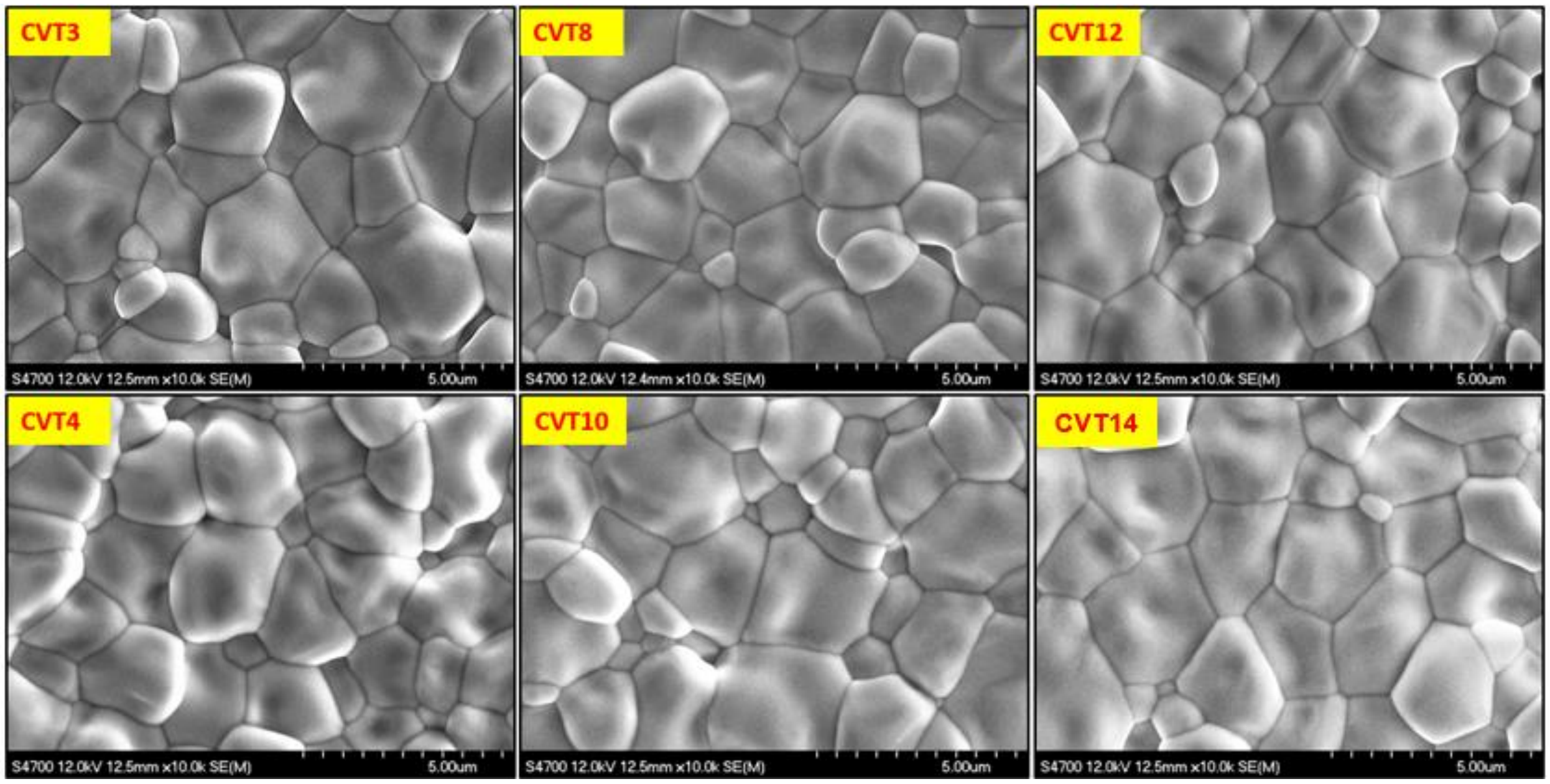

Fig. 16. SEM images from different cutouts representing different field dependence of average surface resistance.

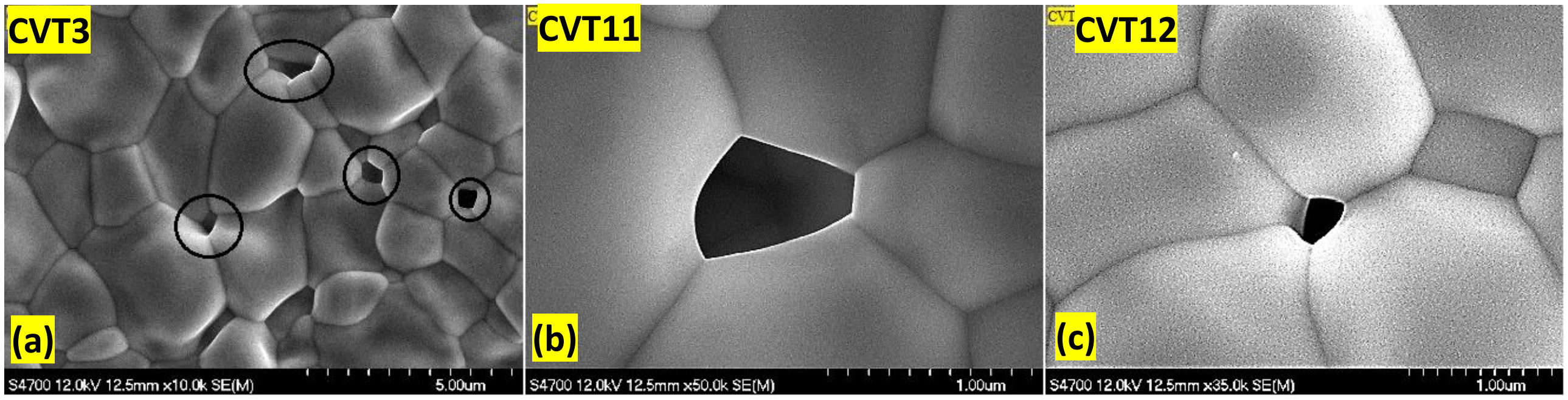

Fig. 17. Several voids (enclosed by ovals) from sample CVT3 are shown in the SEM image to the left. The other two images show pits found in CVT11 (center) and CVT12 (right). Bright outlines indicate sharp edges. Note that the diameter of the pit in CVT11 is about 600 nm.

The microstructure of samples extracted from the top and bottom beam pipes appeared different from those obtained from cavity regions, as shown in Fig. 18. The grain size in the beam pipe cutouts is smaller than that of the cutouts from the cell. No voids were observed in beam pipe cutouts. Local composition of coatings from beam pipe cutouts was similar to that of cutouts from cavity region. Note that the cavity material had higher RRR (≈ 300), while beam pipes were made out of reactor grade niobium (RRR ≈ 40). The average grain size was determined for cutouts with different surface resistance characteristics by counting the number of grains in several SEM images. The estimated average grain size for cavity cutouts was (4.07 ± 0.04) $\mu m^2$ per grain, which appeared significantly larger than those for beam pipes cutouts, see Table *I*. Furthermore, the average grain size for the top beam pipe, 1.35 $\mu m^2$ per grain was smaller than that of the bottom beam pipe, 2.16 $\mu m^2$ per grain.

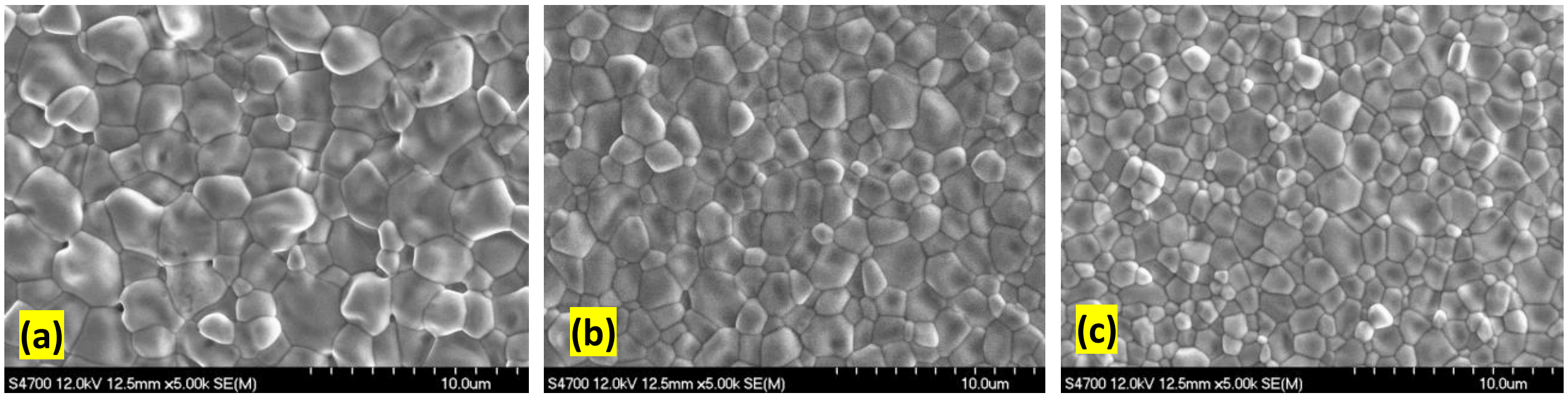

Fig. 18. SEM images obtained from CVT10 (a), bottom beam pipe (b) and top beam pipe (c).

"Patches" with larger irregular grain structures were found in CVT2, CVT4, CVT8, CVT12, and CVT14 and more rarely in beam pipe cutouts. These features were not very common. The size of individual patches varied from a few tens to a few hundred square microns. Small patches were seen at CVT2, CVT4, and beam pipes. Most of those areas had the size that was equivalent to the area covered by a few regular grains, as shown in Fig. 19 (right). Larger patches with the size of a few hundred square microns, as shown in Fig. 19 (left) were encountered in CVT12 and CVT14. Note that these samples showed more coverage of patches than any other cutouts. A few large patches were also observed in CVT8, mostly localized in an unusual area, which will be discussed later. EDS measurement on these smooth areas shows slightly less tin (≈ 21 atomic percent tin) than regular areas (≈ 24 atomic percent tin), a possible indication of a thinner coating. EBSD examination of similar surface obtained in coupon samples revealed that these patchy areas include large single crystalline grains of $Nb_3Sn$ [25]. Finding similar patches in $Nb_3Sn$ coated cavity was recently reported by other researchers [11, 28].

TABLE I. Summary of estimated materials and RF parameters for different cutouts.

| Sample # | Distance from equator(cm) | Sn (at. %) | Void count (per 1000 μm$^2$) | Patch coverage (%) | Average grain size (μm) | Average thickness (μm) | Max. $R_s$ (μΩ) | $R_s$ trend |
|---|---|---|---|---|---|---|---|---|
| 2 | 4.8 | 23.35 | NA | 0.09 | NA | 2.42±0.30 | 0.31 | Field independent |
| 4 | 4.2 | 23.66 | 8 | 0.06 | 4.10 | NA | 0.10 | |
| 8 | 0.6 | 23.69 | 12 | 0.34 | NA | NA | 1.34 | Field dependent |
| 10 | 5.4 | 23.34 | 9 | - | 4.02 | 2.67±0.34 | 4.33 | |
| 12 | 0 | 23.62 | 11 | 0.5 | NA | NA | 3.95 | Field dependent with switch at 4.5 MV/m |
| 14 | 0.6 | 23.56 | 12 | 0.29 | 4.08 | 2.58±0.44 | 3.59 | |
| BP Bottom | 10 | 23.54 | 0 | NA | 2.16 | 2.37±0.19 | | |
| BP top | 10 | 24.53 | 0 | NA | 1.35 | 2.03±0.26 | | |

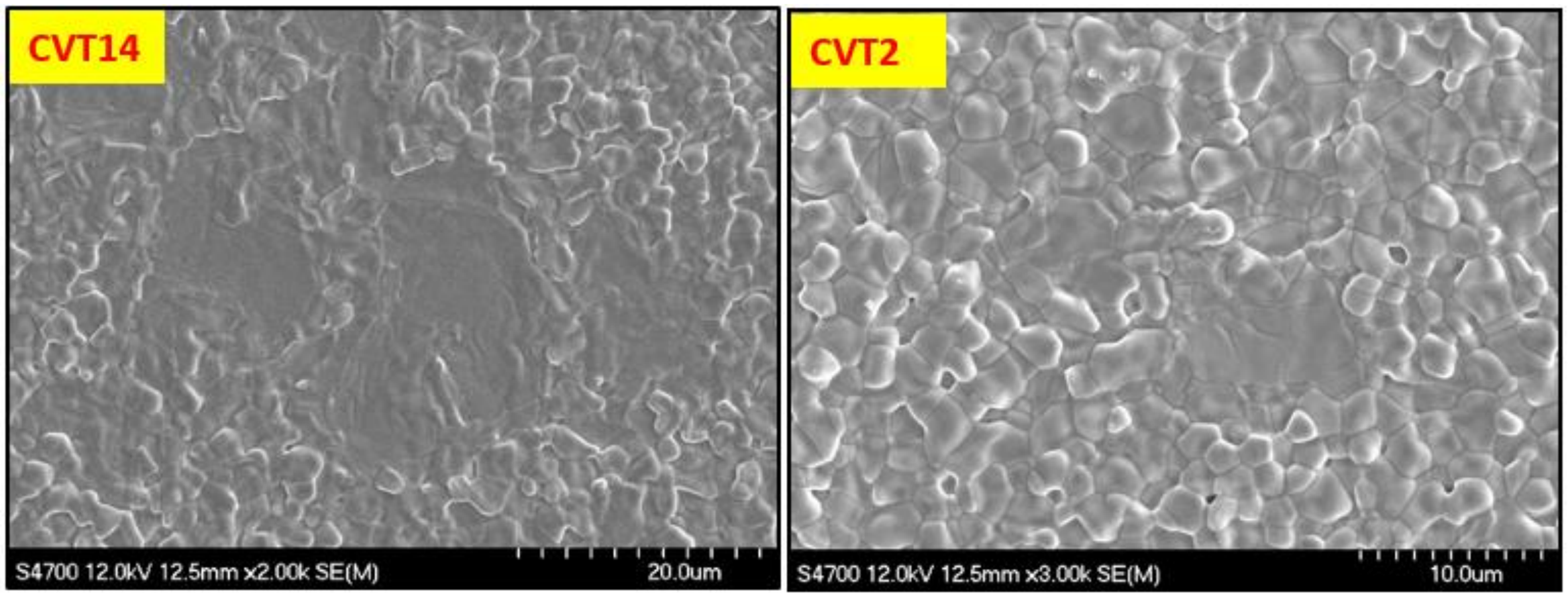

Fig. 19.SEM image of patchy areas obtained from CVT14 (left) and CVT2 (right).

Patches were mostly observed close to niobium grain boundaries. Such patches are often large or appear to have many small patches close together, as shown in Fig. 20. In many cases, the occurrences of patches run parallel to a specific niobium grain boundary for several hundred microns. This was observed in CVT12 and CVT14. SEM images with a magnification of 300-500 were captured from several randomly selected areas of cutouts. The images were then used to locate and measure the size of patchy regions. Table *I* shows the estimated average area covered by patchy regions for each cavity cutout. Patchy regions contribute less than 0.5 % surface coverage in any cutout.

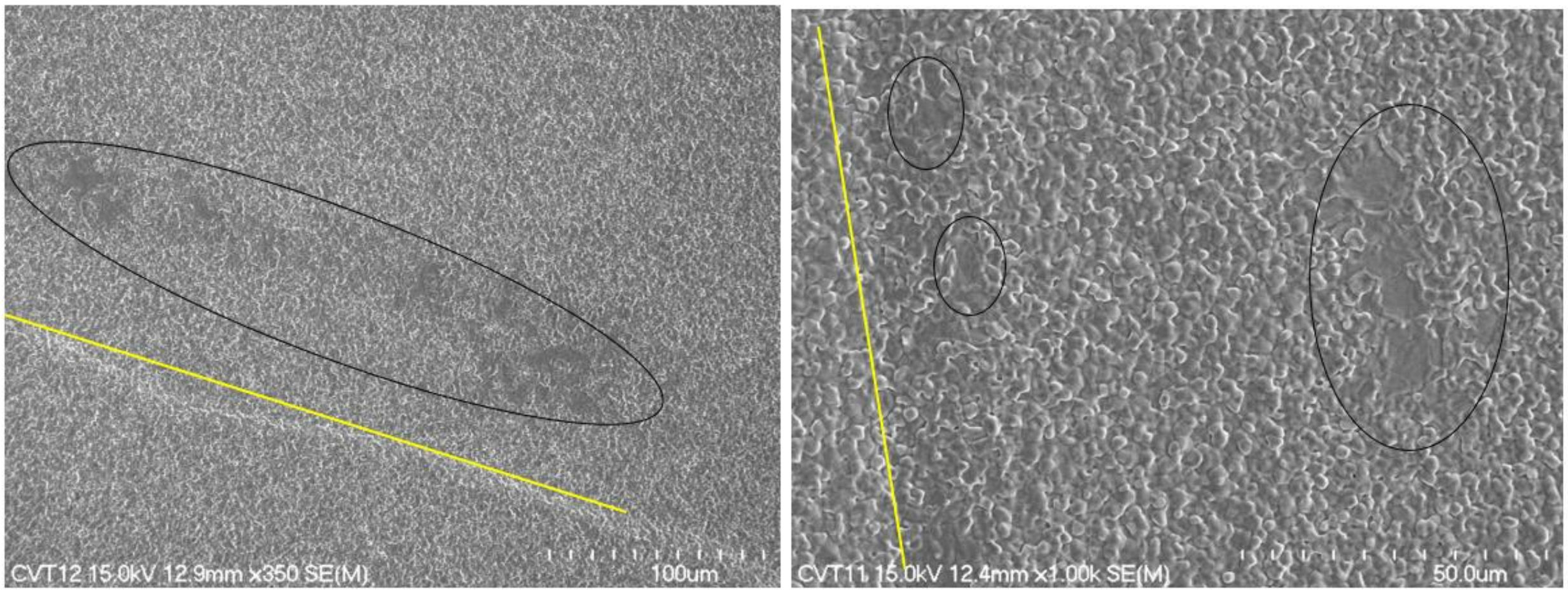

Fig. 20. Patchy areas near grain boundaries. Yellow lines indicate the grain boundaries. Patchy regions are marked with ellipses.

## *B. Topography*

AFM was successfully used only on a few samples due to the curvature of the cutouts. AFM results were consistent with SEM/EDS results to locate the pits and irregular grain structure in the coating. A typical illustration of coating topography on the cavity is shown in Fig. 21. The measurement was carried out in tapping mode using silicon tips with a resonant frequency of 300 kHz, a force constant of 40 N/m, and a tip radius of < 10 nm. Roughness was also measured from 5 $\mu$m x 5 $\mu$m and 50 $\mu$m x 50 $\mu$m scans. Four samples were scanned and average power spectral densities (PSD) were calculated for comparison. The surface height PSD was calculated by Fourier transform of the scan data as in [26, 27]. Note that the area under the PSD curve corresponds directly to root mean square (RMS) roughness. The difference of roughness was not found to be very significant, as shown in Fig. 22. AFM was attempted to obtain the depth profile of a pit. Pits were located in the AFM scans, and depth profile along a line was extracted from different directions as shown in Fig. 23. It has a very different profile than evident in Fig. 17, which can also be an artifact because of sharper edges of the void than the profiling tip. It was found that the depth of these pits varies between 0.5 $\mu$m to 1 $\mu$m.

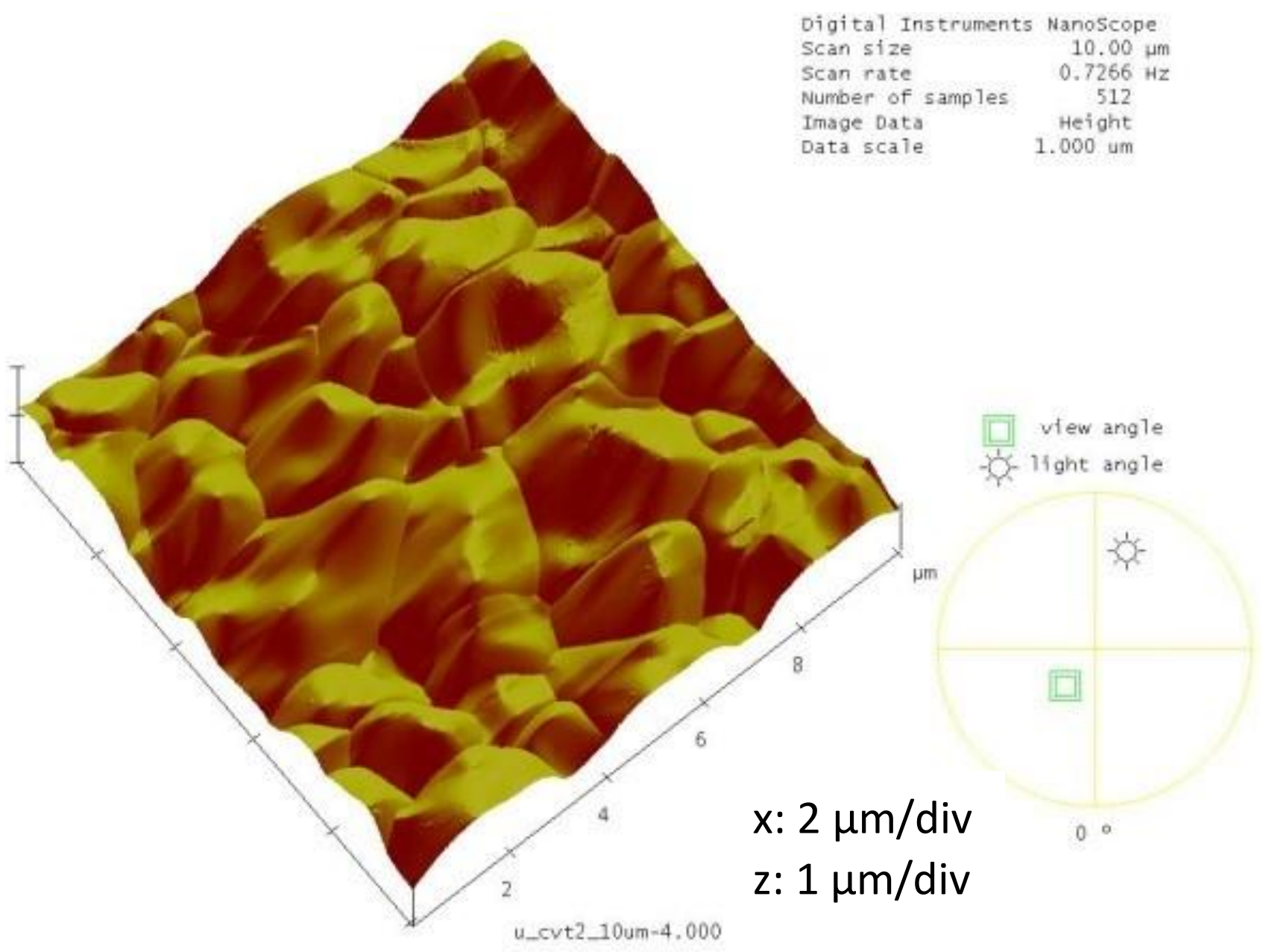

Fig. 21. AFM image of regular $Nb_3Sn$ coatings on CVT2. Roughness is evident with curved convex facets.

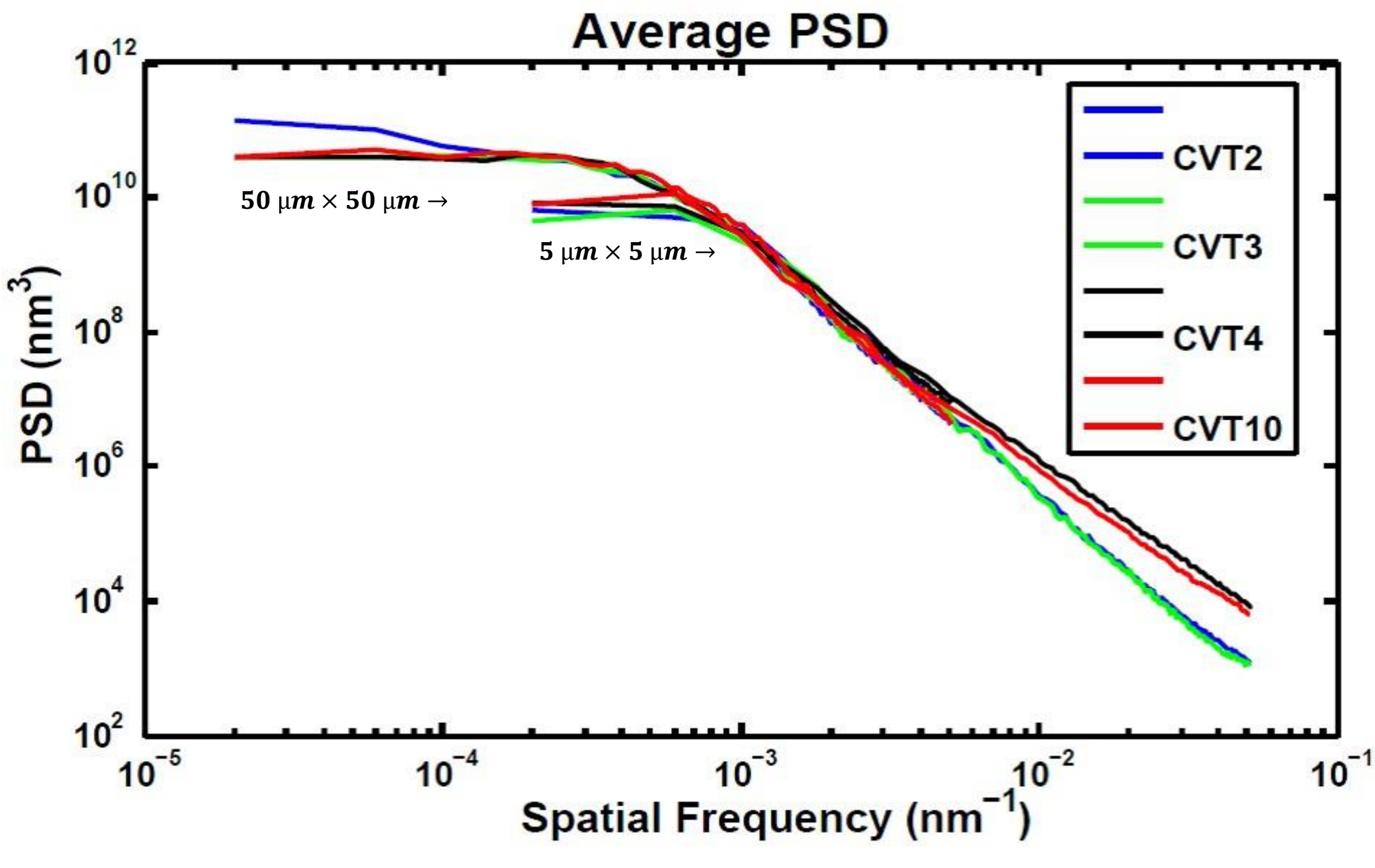

Fig. 22. Comparison of average PSDs of cutouts calculated using AFM data. CVT4 and CVT10 were rougher in the high frequency domain (lateral scale 0.1 $\mu$m and smaller) than CVT2 and CVT3.

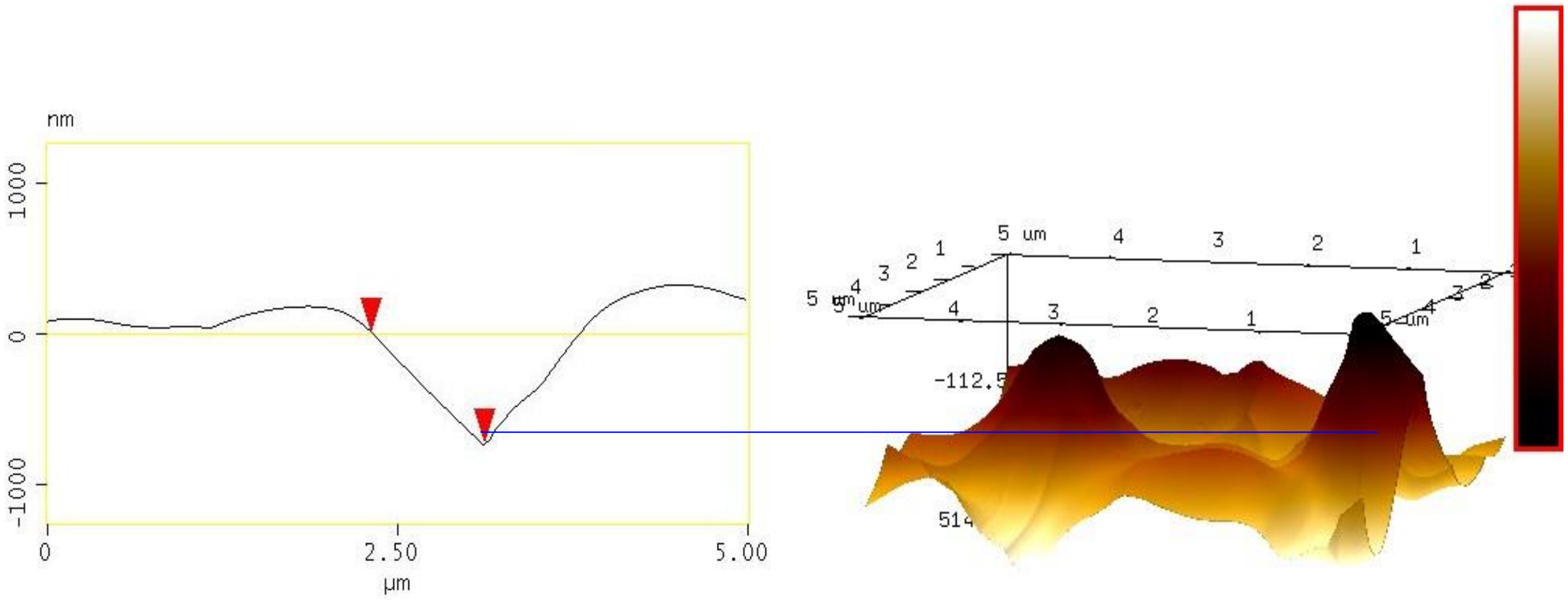

Fig. 23. Profile of a pit observed in CVT2. Pit appears to be ≈ 750 nm deep.

## *C. Cross-section Analysis*

Cross-sections from samples were prepared using a focused ion beam (FIB) after applying a protective layer of platinum onto the coating surface. The obtained cross sections were further polished using an ion mill. SEM images of cross-sections were then captured to measure the coating thickness, as shown in Fig. 24[left]. Coating thickness was measured at several locations of the cutout samples obtained from the cavity region and beam pipes. The coating thickness was found to be greater in cavity regions compared to beam pipes. The coating appeared to be thicker in the bottom beam pipe compared to that in the top beam pipe. The thickness of the coating was found to be as little as 400 nm in the patchy area, see Fig. 24[right], confirming thin coating on such areas.

EBSD images were captured from cutout cross-sections to further examine the structure of the coating. Representative orientation image maps (OIM) obtained are shown in Fig. 25. Columnar grains were normally observed, going from the surface at the top to the $Nb_3Sn$-Nb interface at the bottom. We also observed the formation of non-columnar small grains at the $Nb_3Sn$-Nb interface.

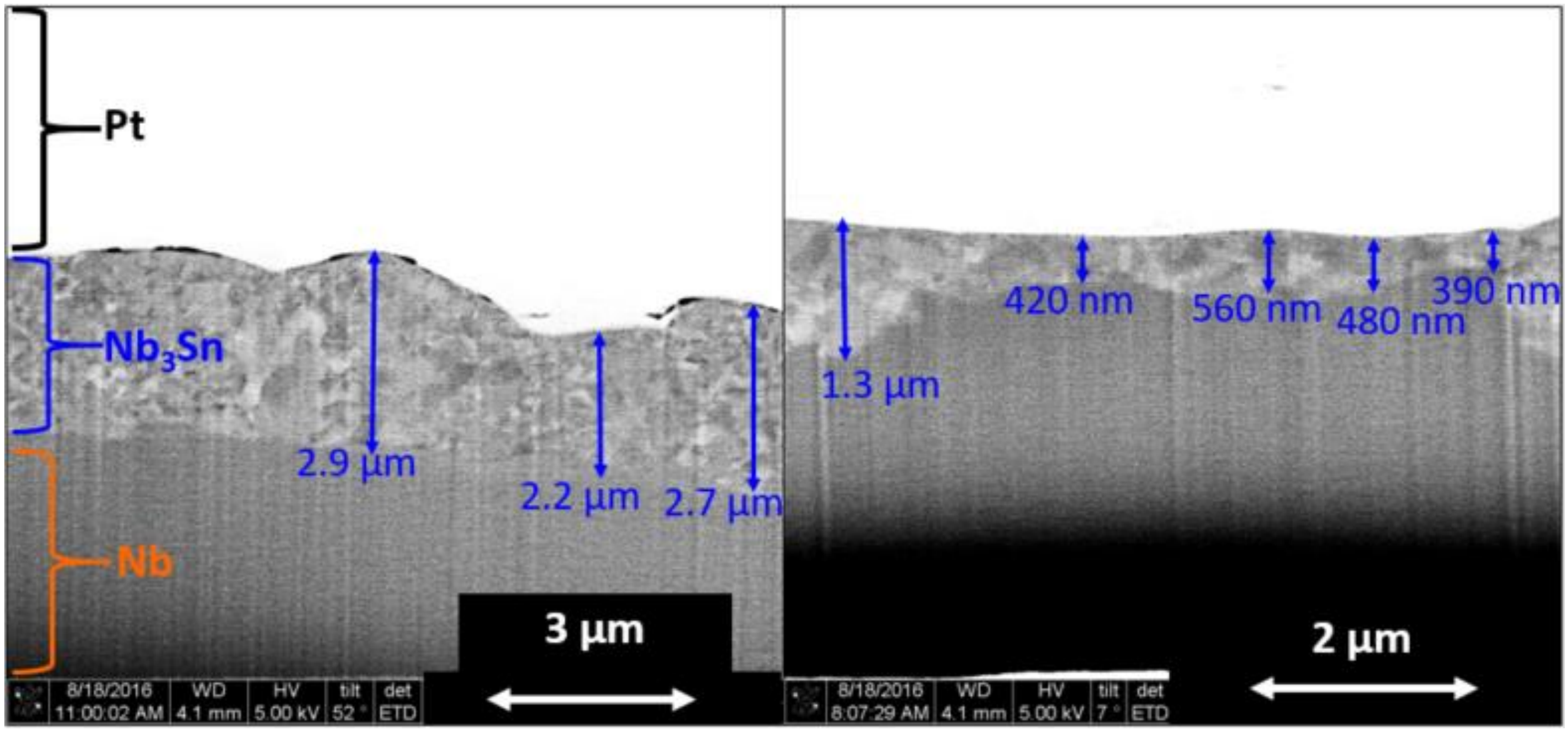

Fig. 24. SEM image of a FIB cross-section obtained from CVT 10 on the left. Thickness varied from 2.34-2.96 $\mu$m. FIB cross-section of a patchy area similar to Fig. 5 on the right. The coating is significantly thin (400-600 nm) compared to the neighboring area (> 1.3 $\mu$m).

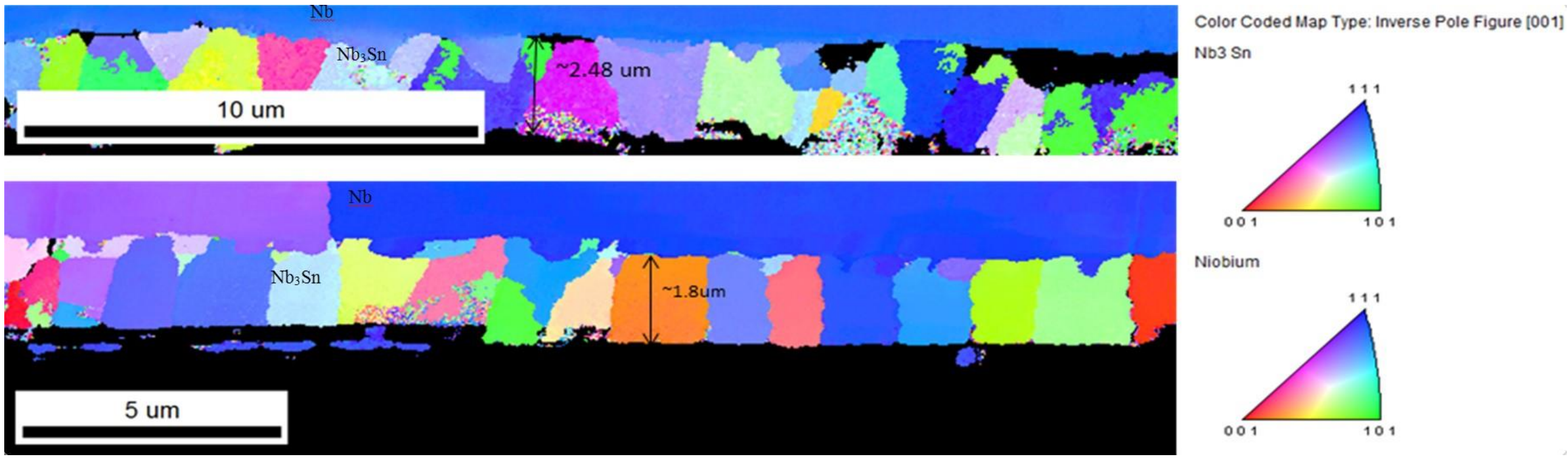

Fig. 25. EBSD image from sample CVT 10 at the top which is cut from the cavity region. The image at the bottom is from the beam pipe cutout. Beam pipe has a thinner coating with more columnar grains than CVT10. Some instrumental artifacts (dark area in between Nb-$Nb_3Sn$ grains) are present.

## *D. Other Defects*

Besides patches and pits, other defects were found on some samples. One such defect is bright spots observed on coated surfaces by SEM, as shown in Fig. 26. The contrast difference might have arisen due to the difference in topography. These features were found sometimes on CVT2 and CVT4, but frequently seen in CVT10. These spots were found to have different shapes with carbon-enriched boundaries as shown in Fig. 27. Note that the ratio of Nb to Sn was not altered within these bright spots when examined with EDS, suggesting that the SEM signature is due to very thin surface contamination. Unlike other cutouts, CVT10 had a large area with a different appearance than usual coating as shown in Fig. 28. It comprises the usual $Nb_3Sn$ grains, but dark spots appear in each of those grains. Similar dark areas were also noticed in intergranular spots. EDS found an excess of carbon and oxygen from dark spots.

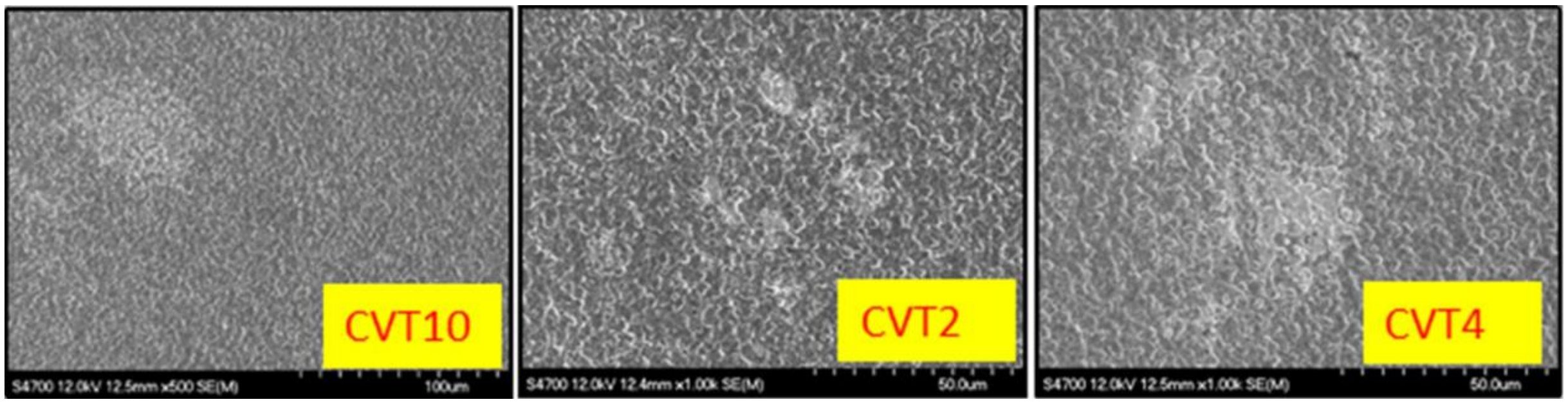

Fig. 26. Bright spots observed in different cutouts.

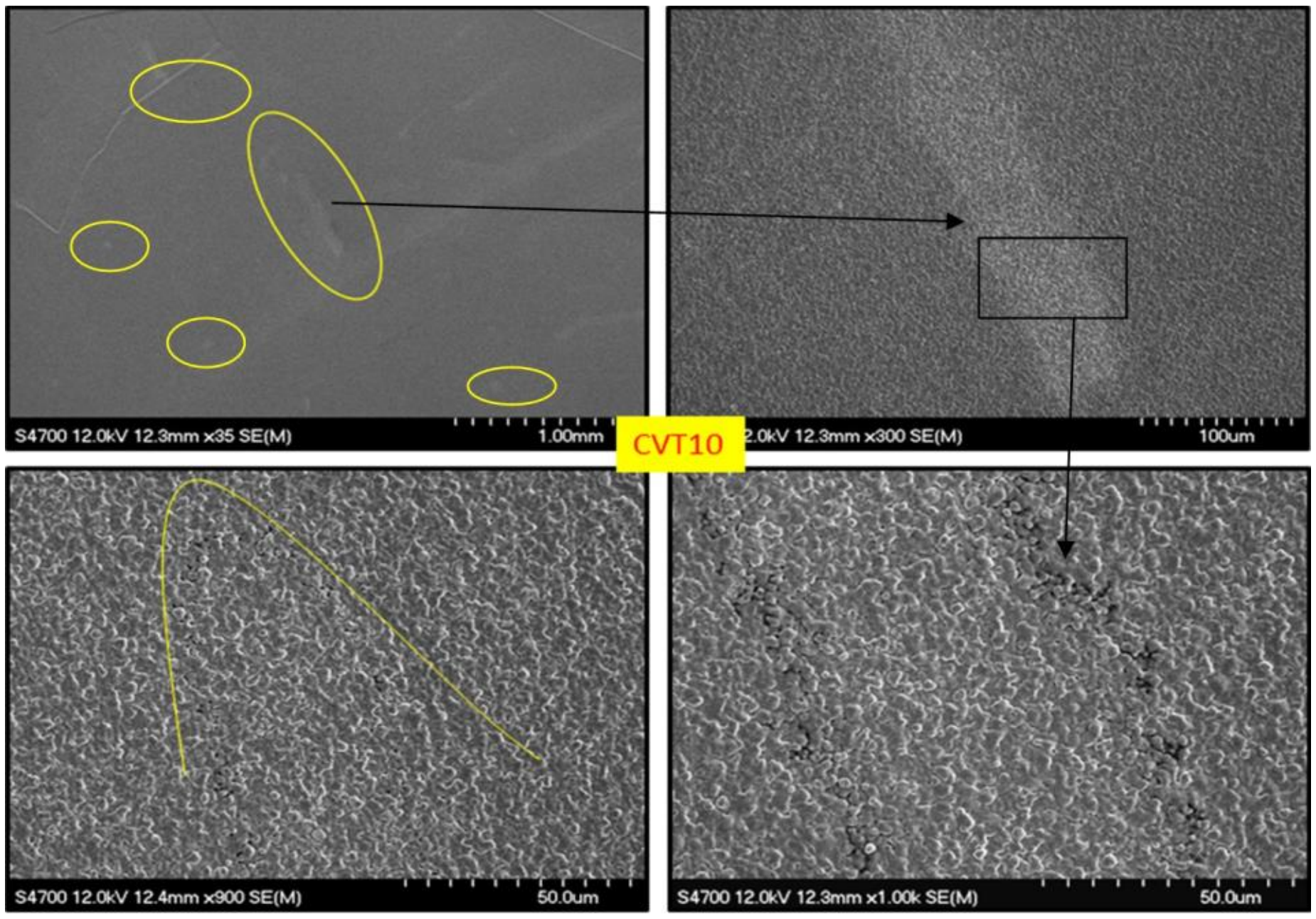

Fig. 27. Bright spots observed in CVT10. Note that the dark boundaries of these features show a clear presence of carbon.

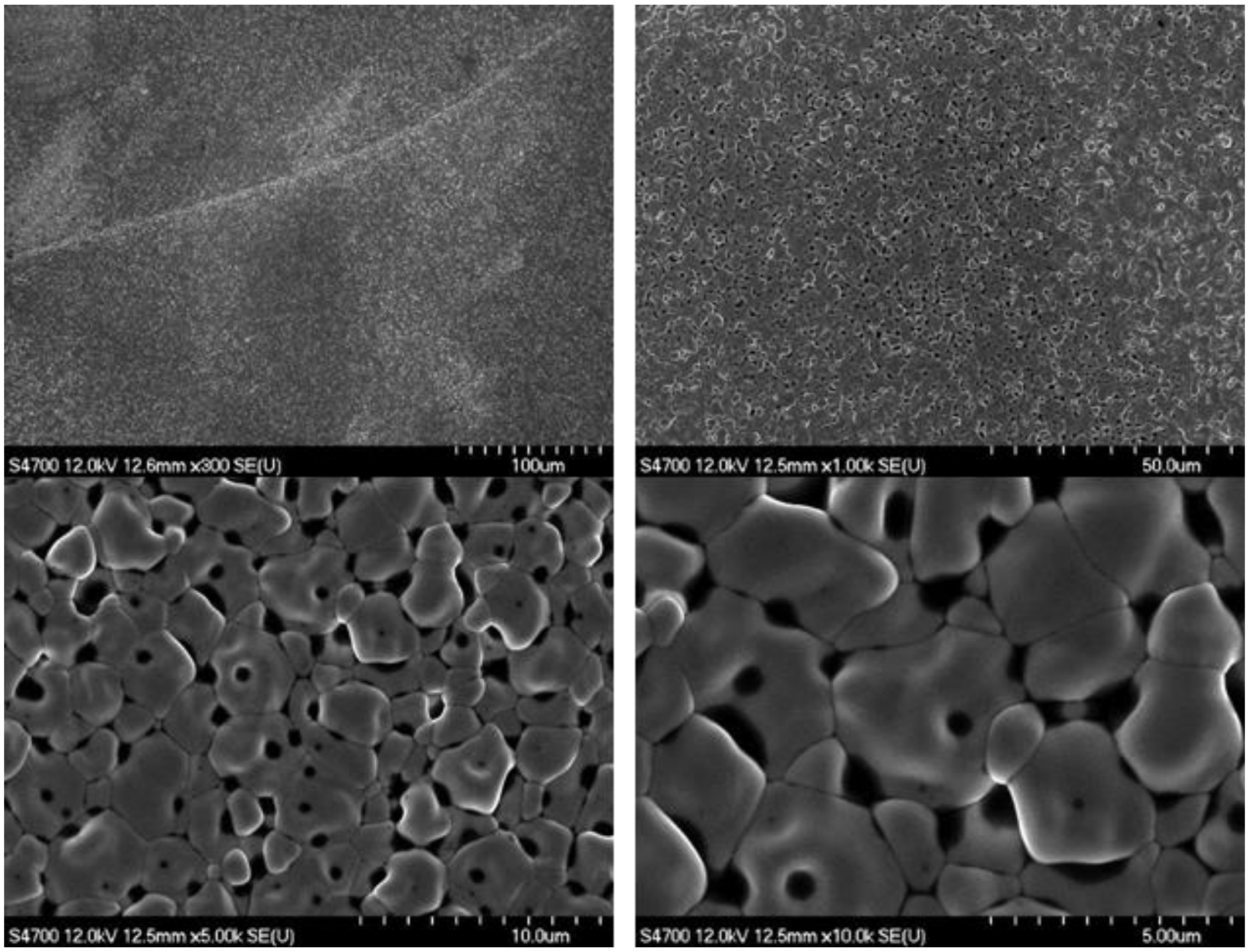

Fig. 28. An unusual appearance of coating in CVT10.

An AFM image collected from CVT10, see Fig. 29, also shows the presence of residues on the surface. A circular feature of 100um diameter was present in CVT14 as shown in Fig. 30. The circular area was found recessed compared to its adjacent areas, considering the contrast in the SEM image. The edge of this feature showed a significant presence of carbon and oxygen. The composition and grain structure were found similar inside and outside of this feature. Another noticeable area was discovered in CVT8. The area, which appeared relatively dark in SEM images is shown in Fig. 31(a). Note that the area covered by this constituent was more than 40000 $\mu m^2$, that is, the diameter was ≈ 700 $\mu$m. SEM images show well defined boundaries, see Fig. 31(b, c, and e) that distinguish this area from areas with usual looking coating. Large patches were also seen at the boundary, as shown in Fig 31(f). Fig 31(d) and 31(e) respectively show grains from the unusual area under discussion, and nearby area with usual appearances. Corresponding EDS spectra show significant amounts of carbon from dark areas compared to regular areas. This unusual area of CVT8 has similar characteristics of the unusual area found in CVT10, Fig. 28.

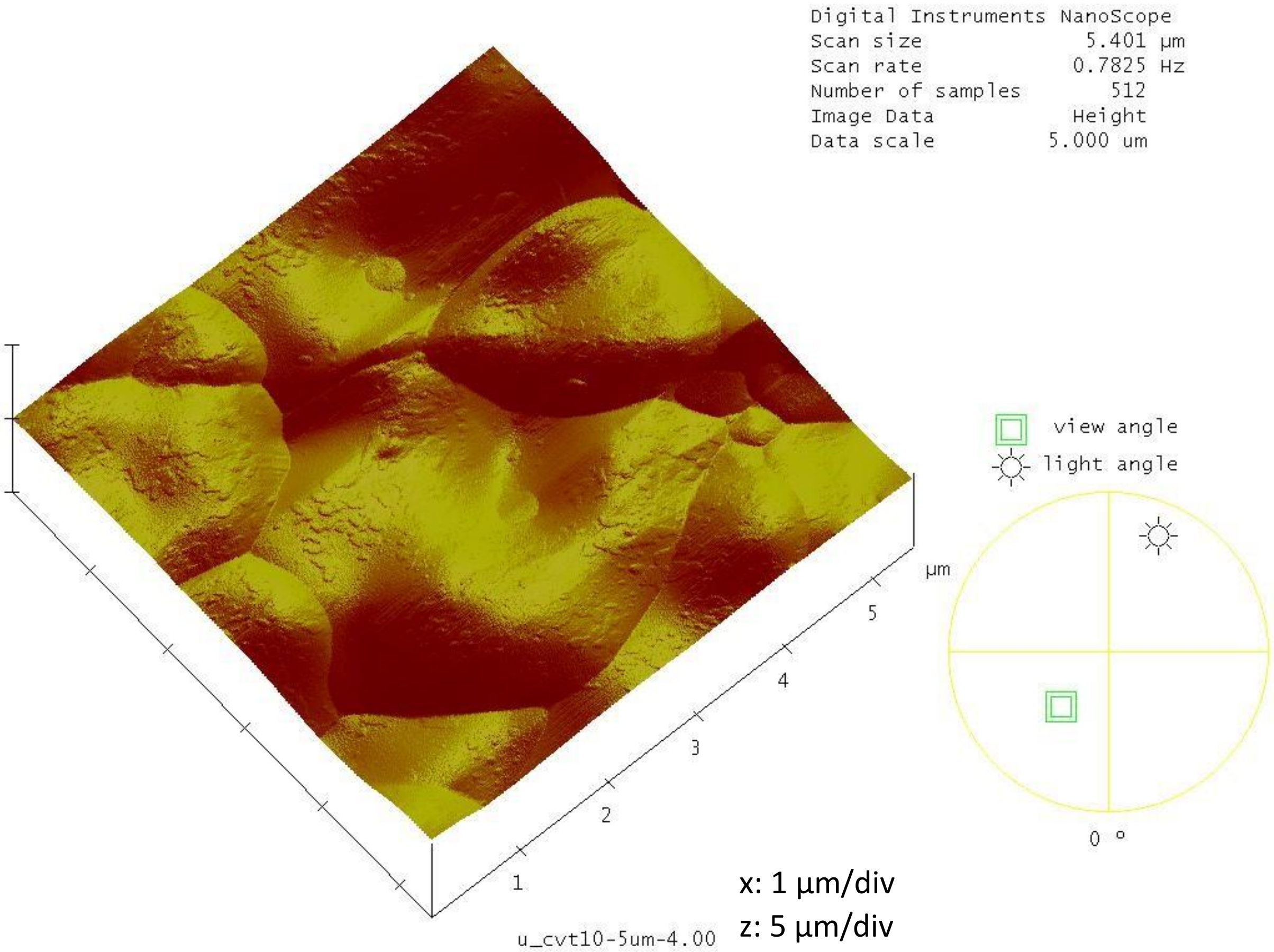

Fig. 29. AFM image from CVT 10. Note the residue covering the surface.

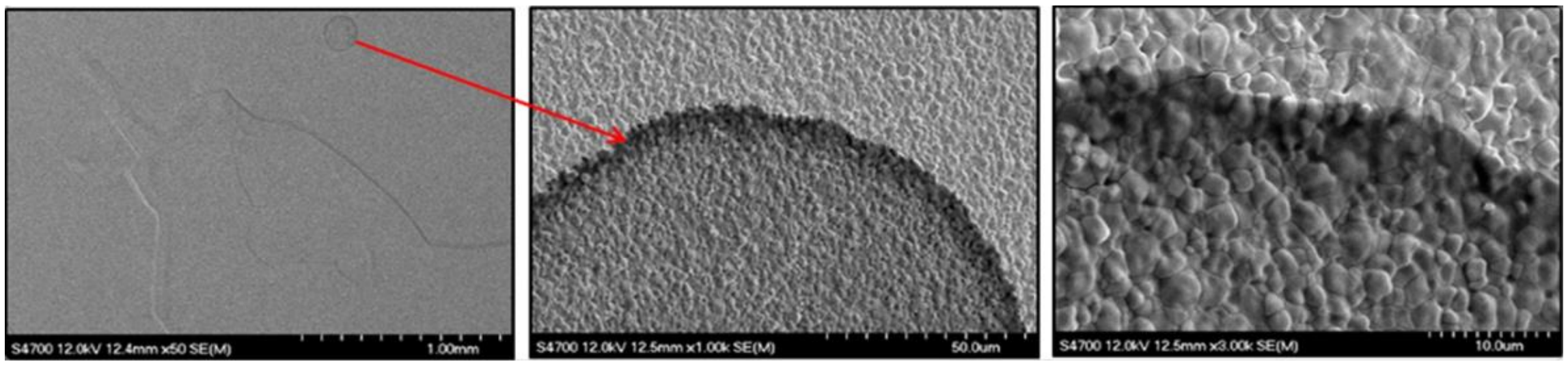

Fig. 30. Defect found in CVT14.

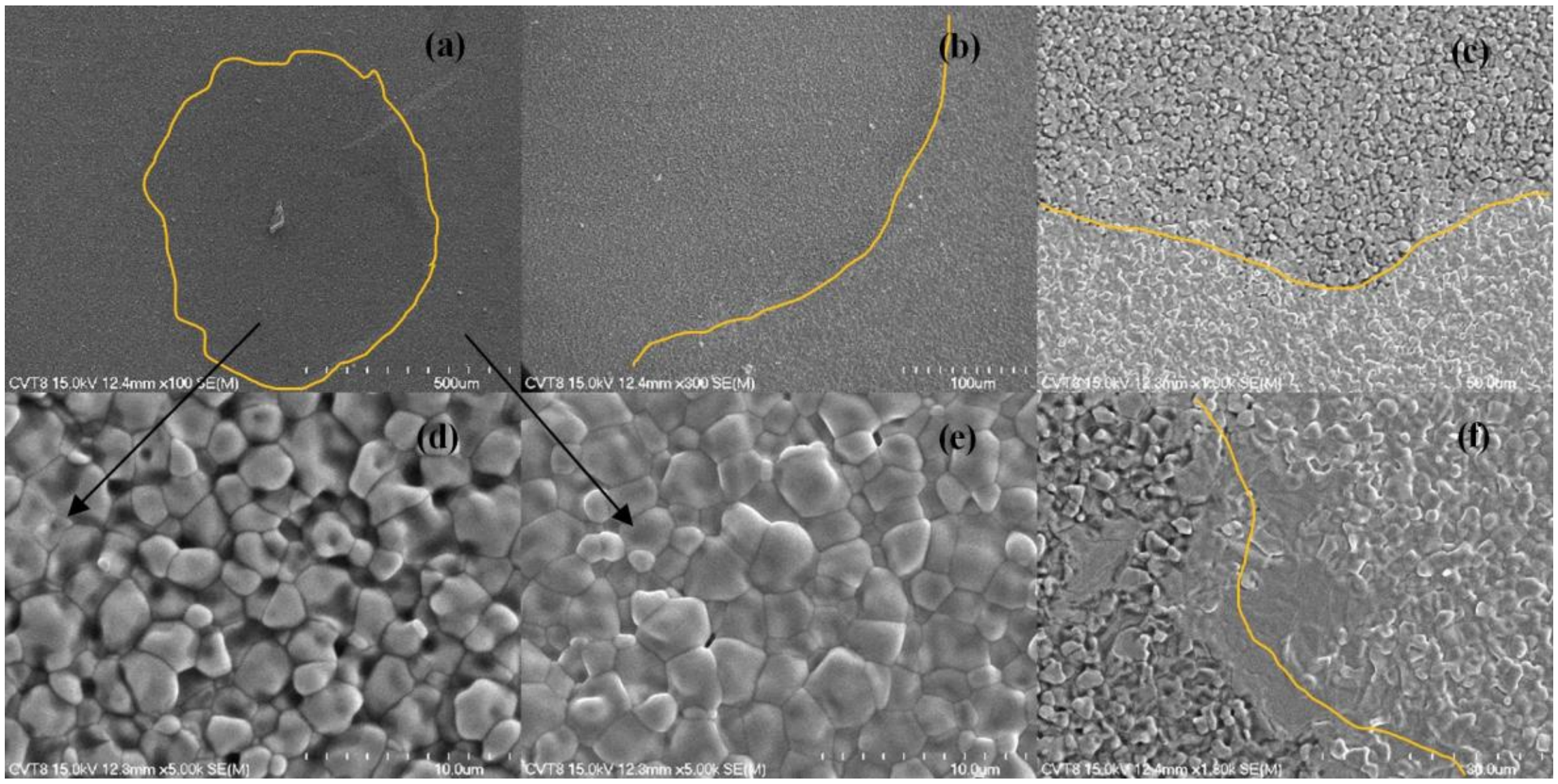

Fig. 31. Detailed structure within the defect found in CVT8. (a) shows the unusual area marked with yellow lines. (b), (c) and (e) shows the transition from unusual area to regular (c). (d) and (f) show close-up views of grains from unusual and regular areas respectively. Note the patchy region found close to the transition.

# VIII. DISCUSSION

Sample analysis and RF measured transition temperature both showed good coverage of the $Nb_3Sn$ coating in the cavity. Patchy areas containing irregular grains were present in several cutouts. Cross-section measurements revealed that such irregular $Nb_3Sn$ grains are significantly thinner than the grain in regular regions, Fig. 24. Since RF field penetration in $Nb_3Sn$ is in the order of a few hundred nanometers, the thickness of such patchy areas is not large enough to fully shield the underlying $Nb_3Sn$-Nb interface and underlying niobium from the RF field. Underlying niobium has higher surface resistance and will cause additional RF losses compared to the regular $Nb_3Sn$ grains. A correlation between the relative abundance of the irregular grains and stronger RF field dependence is seen in the thermometry data. Thermometry data for the CVT8, CVT12, CVT14 cutouts indicates average surface resistance above 1 $\mu\Omega$ at $E_{acc}$ = 6 MV/m and a field dependence exponent above 3, and the relative irregular single crystal density in these samples exceeds 0.2 %. Inefficient RF shielding by the irregular single crystal grains contributes to the additional RF loss in these samples. The mechanisms that lead to the formation of such areas are not clear yet. One observation is that all three cutouts CVT8, CVT12, and CVT14 are from the equator region of the cavity, which seems to be more likely to develop such features, Fig. 32.

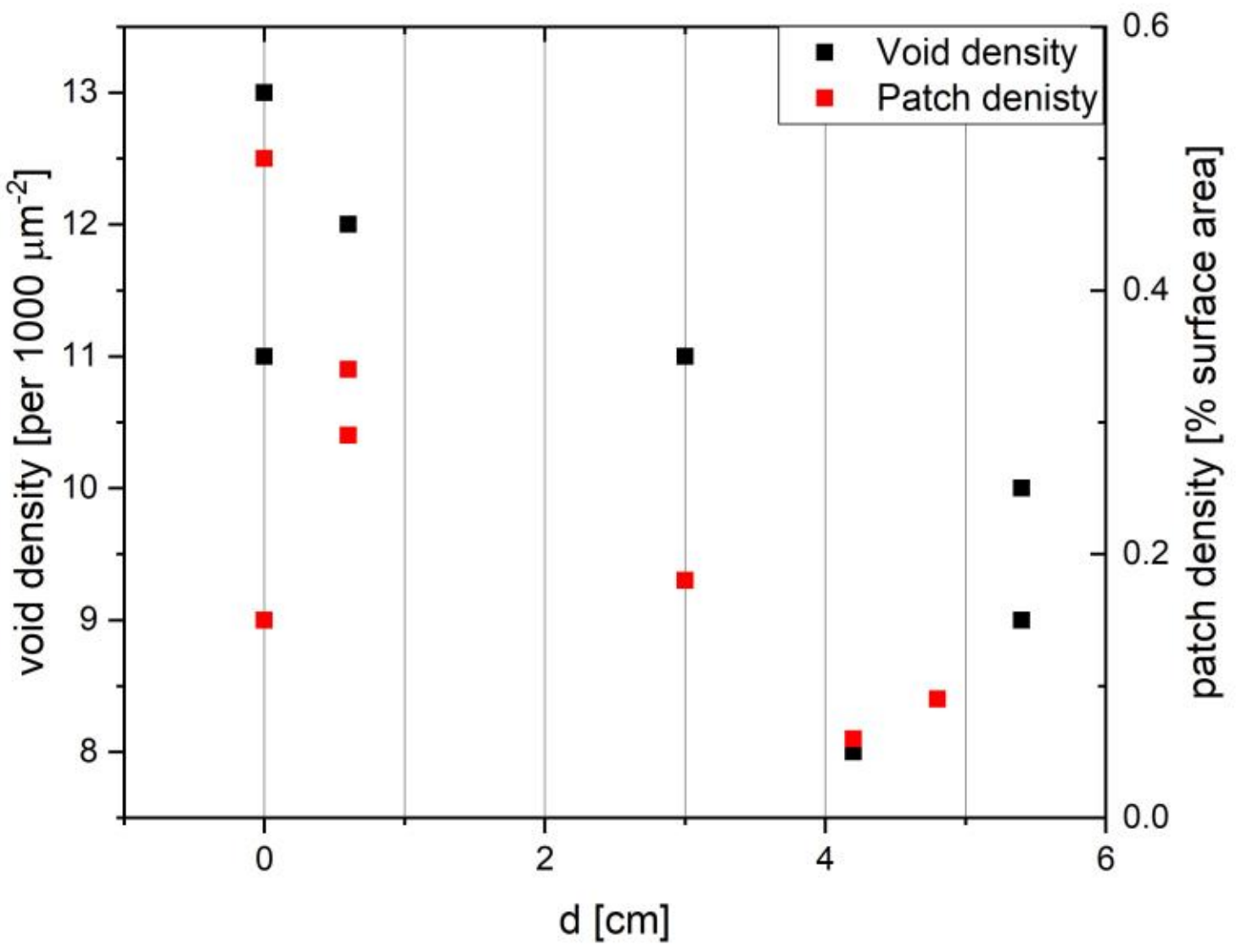

Fig. 32. Defect density as a function of distance from the equator. Note the negative correlation, i.e., defects are more likely to occur in the regions close to the equator.

Electron beam welding areas feature larger grains that form during the solidification process of the weld seam. The cavity equator region is also the region, which is the furthermost out-of-sight location from the tin evaporation source. Since tin flux to such regions is supplied via scattering and surface diffusion, the equator regions may see less tin flux than the in-sight areas, and the formation of thin regions may be linked to low tin flux. With less tin supplied to these regions, one expects overall less tin content and hence a thinner $Nb_3Sn$ layer thickness. Thickness measurements of the $Nb_3Sn$ films in different cutouts did not show significant variation with the cutout position on the cavity surface. The films both on in-sight and out-of-sight surfaces have similar thicknesses, Table *I*. This measurement indicates that the amount of tin available for grain boundary diffusion and $Nb_3Sn$ formation during the coating growth is similar or exceeds the diffusion rate for both in-sight and out-of-sight cavity regions during the film coating process. The amount of available tin in the out-of-sight areas will depend on the sticking coefficient and surface ad-atom mobility. $Nb_3Sn$ growth is done typically at temperatures above 1000 °C in the vapor diffusion process, where the sticking coefficient can be lower, and the surface ad-atom mobility is higher than those at the nucleation temperature [34, 35], helping coating uniformity. The preceding nucleation process is done at about 500 °C, where the sticking coefficient is higher, and surface ad-atom mobility is lower, impeding uniformity of tin coating. During the nucleation stage, some out-of-sight areas may remain poorly covered with tin. Once the process progresses to the growth stage, such areas will present large bare niobium areas to arriving tin atoms with higher mobility at this stage. Due to the absence of pre-nucleated centers and high tin mobility, such areas will promote the growth of large single crystal grains, which will subsequently grow slower due to the absence of grain boundaries. This model indicates that film nucleation is critical to growing film without thin regions, and the film grown on cavity C3C4 may have suffered from low tin flux during the nucleation phase.

In the past, researchers also linked such thin areas to the substrate grain orientation [8]. They reported that single crystal substrate with (111) and (531) planes resulted in more patchy areas than the substrates with (110) and (100) orientation. Beam welding areas are likely to have large grains, on the order of 100 $\mu$m, resembling single crystal substrates, which may include similar crystal orientations favorable for non-uniformity. It has been reported recently that the accumulation of tin particles during the nucleation stage can vary for different grains of niobium [33]. [37] has reported finding such areas in a large grain sample [18]. To solve this problem, the substrate pre-anodization was employed. The pre-anodization technique was recently utilized at Cornell University to overcome these patchy regions [29]. Sample studies at Jefferson Lab also indicated the formation of patch-free coating in coupon samples with small amounts of tin [30]. The average coverage of patchy regions in our dissected cavity (< 0.25%), which was not pre-anodized, is very similar to the result obtained with pre-anodizing by researchers at Cornell University. Note that the nucleation process used at Jefferson Lab (1 hour at 500 ˚C with 3 mg/cm$^2$ of $SnCl_2$) differs from that of the coating protocol used at Cornell (5 hours at 500 ˚C with 10 $\mu$g/cm$^2$ of $SnCl_2$). It is possible that the larger amount of $SnCl_2$ may be expected to provide higher Sn vapor pressure to nucleate uniformly the large surface area of a cavity. Also, the coating temperature of $Nb_3Sn$ at Jefferson Lab, 1200 ˚C, is higher than that used at Cornell, 1100 ˚C, providing higher flux of tin during coating. It is proposed that patches have thinner coating because of fewer grain boundaries and longer grain boundary diffusion paths to supply fresh Sn to mid-grain at the Nb-$Nb_3Sn$ interface. Grain boundary diffusion is understood to be the primary mode of tin transportation to $Nb_3Sn$-Nb interface for coating growth [36].

AFM images revealed depressions on $Nb_3Sn$ grains formed by curved facets. These depressions were also visible in SEM images as a dark spot in the middle of $Nb_3Sn$ grains. These depressions in many instances appear to be associated with local contamination, as seen in Fig. 29. The average RMS roughness was found close to 100 nm in the cutout samples for 5 $\mu$m x 5 $\mu$m scans, slightly higher than 70 nm in samples coated before [31]. The difference is expected because of relatively larger grains and thicker coating in cavity cutouts [30]. The variation of roughness between examined samples was not very significant when average PSDs from the data obtained with AFM were compared.

Microscopic pits were observed in all the cutouts except those from beam pipes. Roughness and pits appeared not to be responsible for the dominant localized heating during RF testing. These may affect the cavity performance globally and with increasing field. Grain growth competition and rapid coalescence of small grains during the deposition step are speculated to create such structures.

Coating thickness and grain size were found to vary between the cavity region and beam pipes. One obvious difference between them was the RRR value of the fabrication material. There has been limited research to understand the relationship between RRR of substrate niobium and $Nb_3Sn$ growth process. Peiniger et al. reported that the density of $Nb_3Sn$ nucleation centers was strongly reduced if one uses medium RRR niobium (RRR = 120) compared to low RRR niobium (RRR ≈ 40) [32]. The phenomenon causing the differences has not yet been established. Considering the potential for more nucleation sites in beam pipe (low RRR niobium) compared to cavity cell areas (high RRR), the observed difference in the number of grains or the average grain size may be understood. Grain size variation between cavity cutouts was not very significant, but the difference in grain size between the top beam pipe and bottom beam pipe was significant. One possible reason for such a difference could be the distance from the tin source during the coating. The top beam pipe, which is farther from tin source, may have received a lower flux of tin during the coating process compared to the bottom. Since we did not see such an asymmetry between cavity cutouts, another potential reason for smaller grains in the top beam pipe could be the proximity to the top heat shield and pump line. The temperature of the top beam pipe during the coating might have been lower than the beam pipe if the heat shield was not perfect. Note that we have seen condensed tin on niobium foil covering the top beam pipe during the coating.

Other carbon-enriched defects, similar to those depicted in Fig.s 28 and 27 indicate that some impurity was already present at these spots on the starting substrate niobium. If that is the case, it appears that impurity segregation to the surface can happen without a dramatic influence on $Nb_3Sn$ growth. Unusual and big carbon enriched areas present in CVT8 and CVT10 are strong candidates for localized heating. The circular defect found in CVT14, Fig. 30, could be another similar case where the niobium might have a circular depression, susceptible to retaining carbonaceous impurities prior to the coating. Such defects might be present around the equator as welding artifacts. Residue, observed in CVT10 with AFM, Fig. 29, may not appear in SEM image because of the limited sensitivity of SEM/EDS, but It could be something that can also impair the RF performance. The higher density of patchy regions in CVT12 and CVT 14 appear to be well correlated with the observed surface resistance switch at ~ 4.5 MV/m. The carbon-rich features seem to correlate with strong field dependent resistance. The absence of such features results in a material with relatively field-independent surface resistance.

Since intermetallic A15 compounds are extremely brittle, $Nb_3Sn$ is vulnerable to fracture. It is known that brittle fractures can propagate faster along the direction perpendicular to the applied stress. Cracks observed in cutouts typically started from the edge and propagated in the same direction away from the edge, indicating that they were formed due to the applied mechanical stress during dissection.

## IX. CONCLUSION

Several areas with different RF loss characteristics were located with thermometry mapping measurements during the RF test of a $Nb_3Sn$-coated cavity. Samples from these areas were extracted and analyzed with different materials characterization techniques. RF analysis of the cutout regions showed three systematic trends in surface resistance with increasing accelerating field: weak field dependent, strong field dependent and field dependent switch (at 4.5 MV/m). Each cutout had similar microstructures, thickness, and composition of $Nb_3Sn$. Voids were commonly seen in each cutout extracted from the cavity region but not in the samples from the beam pipes, which also had relatively smaller grain sizes. This indicates that the purity of niobium substrate influences the microstructural properties of $Nb_3Sn$. Several cutouts possess patchy regions with thinner coating. A higher density of patchy regions in the equator region seems to result in surface resistance switch at 4.5 MV/m and strong field dependency. The presence of large regions with carbon contamination appears to contribute to strong field dependent surface resistance.

# X. ACKNOWLEDGMENTS

Authored by Jefferson Science Associates, LLC under U.S. DOE Contract No. DE-AC05-06OR23177. This material is based upon work supported by the U.S. Department of Energy, Office of Science, Office of Nuclear Physics. Work at The College of William and Mary was supported by the Office of High Energy Physics under grant SC0014475. We would like to thank Steve Castignola, Kirk Davis, Chris Dreyfuss, Stephen Dutton, Jim Follkie, Danny Forehand, Tom Goodman, Teena Harris, Pete Kushnick, Roland Overtone, and Tony Reilly for technical support; Gigi Ciovati, Pashupati Dhakal, Rongli Geng, Peter Kneisel, Ari Palczewski, Larry Phillips, Josh Spradlin, and Anne-Marie Valente-Feliciano for comments and suggestions during various phases of the project; and Robert Rimmer for continued support. Thanks to Olga Trifimova for her help with AFM analysis at William and Mary applied research center.